%% file: conext17.tex
\documentclass[10pt, sigconf]{acmart}

\renewcommand\footnotetextcopyrightpermission[1]{} 
\usepackage{booktabs} 
\usepackage[tight,footnotesize]{subfigure}
\usepackage{algorithm,algorithmic}
\usepackage{times}
\usepackage{graphics}
\usepackage{graphicx}
\usepackage{epstopdf}

\usepackage{epsfig}
\usepackage{amstext}
\usepackage{amssymb}
\usepackage{amsmath}
\usepackage{url}
\usepackage{psfrag}
\usepackage{verbatim}
\usepackage{color, colortbl}

\newcommand {\bear}{\begin{eqnarray}}
\newcommand {\eear}{\end{eqnarray}}
\newcommand {\bearn}{\begin{eqnarray*}}
\newcommand {\eearn}{\end{eqnarray*}}
\newcommand {\barr}{\begin{array}}
\newcommand {\earr}{\end{array}}

\def\b1{\mathbf{1}}

\definecolor{Gray}{gray}{0.9}
\definecolor{LightCyan}{rgb}{0.88,1,1}

\newcommand{\be}{\begin{equation}}
\newcommand{\ee}{\end{equation}}
\newcommand{\bi}{\begin{itemize}}
\newcommand{\ei}{\end{itemize}}
\def\bearn{\begin{eqnarray*}}
\def\eearn{\end{eqnarray*}}

\newcommand{\hide}[1]{}

\newcommand*\xbar[1]{%
  \hbox{%
    \vbox{%
      \hrule height 0.5pt 
      \kern0.5ex
      \hbox{%
        \kern-0.1em
        \ensuremath{#1}%
        \kern-0.1em
      }%
    }%
  }%
}

\newenvironment{enumerate*}%
  {\begin{enumerate}%
    \setlength{\itemsep}{1pt}%
    \setlength{\parskip}{1pt}}%
  {\end{enumerate}}

\newenvironment{itemize*}%
  {\begin{itemize}%
    \setlength{\itemsep}{0.8pt}%
    \setlength{\parskip}{0.8pt}}%
  {\end{itemize}}

\setcopyright{none}

\acmConference[CoNEXT'17]{The 13th International Conference on emerging Networking EXperiments and Technologies}{2017}{Seoul, South Korea} 
\begin{document}
\title{Multipath IP Routing on End Devices: \\
Motivation, Design, and Performance}

\author{Liyang Sun}
\affiliation{%
  \institution{ECE, NYU}
  \streetaddress{2 MetroTech Center}
  \city{Brooklyn} 
  \state{NY} 
  \postcode{11201}
}
\email{ls3817@nyu.edu}

\author{Guibin Tian}
\affiliation{%
  \institution{ECE, NYU}
  \streetaddress{2 MetroTech Center}
  \city{Brooklyn} 
  \state{NY} 
  \postcode{11201}
}
\email{gbtian@gmail.com}

\author{Guanyu Zhu}
\affiliation{%
  \institution{ECE, NYU}
  \streetaddress{2 MetroTech Center}
  \city{Brooklyn} 
  \state{NY} 
  \postcode{11201}
}
\email{gz623@nyu.edu }

\author{Yong Liu}
\affiliation{%
  \institution{ECE, NYU}
  \streetaddress{2 MetroTech Center}
  \city{Brooklyn} 
  \state{NY} 
  \postcode{11201}
}
\email{yongliu@nyu.edu}

\author{Hang Shi}
\affiliation{%
  \institution{Huawei}
  \streetaddress{2330 Central Expy}
  \city{Santa Clara}
  \state{CA}
  \postcode{95050}
}
\email{Hang.Shi@huawei.com}

\author{David Dai}
\affiliation{%
  \institution{Huawei}
  \streetaddress{2330 Central Expy}
  \city{Santa Clara}
  \state{CA}
  \postcode{95050}
}
\email{david.h.dai@huawei.com }

\begin{abstract}
Most end devices are now equipped with multiple network interfaces. Applications can exploit all available interfaces and benefit from multipath
transmission. Recently Multipath TCP (MPTCP) was proposed to implement multipath transmission at the transport layer and has attracted lots of attention from academia and industry. However, MPTCP only supports TCP-based applications and its multipath routing flexibility is limited.  In this paper, we investigate the possibility of orchestrating multipath transmission from the network layer of end devices, and develop a Multipath IP (MPIP) design consisting of signaling, session and path management, multipath routing, and NAT traversal.  We implement MPIP in Linux and Android kernels. Through controlled lab experiments and Internet experiments, we demonstrate that MPIP can effectively achieve multipath gains at the network layer. It not only supports the legacy TCP and UDP protocols, but also works seamlessly with MPTCP. By facilitating user-defined customized routing, MPIP 
can route traffic from competing applications in a coordinated fashion to maximize the aggregate user Quality-of-Experience. 
\end{abstract}

\maketitle


\input{intro}
\input{design}

\input{evaluation}
\input{related}

\vspace{-1.5mm}
\section{Conclusions and Future Work}
\label{sec:conclusion}
In this paper, we developed MPIP, a complete design of multipath transmission at the network layer of end devices. MPIP consists of signaling, session and path management, multipath routing, and NAT traversal. MPIP can be used by both TCP and UDP-based applications. It also works seamlessly with MPTCP, and supports user-defined routing strategies. We implemented MPIP in Linux and Android kernels. Through extensive lab and Internet experiments, we demonstrated that MPIP can transparently support flexible and coordinated routing for diverse applications to achieve multipath gains. 

MPIP is only our first attempt for implementing multipath transmission at the network layer. The signaling and feedback mechanisms can be further optimized to reduce its overhead and improve its robustness. The delay-based load balancing  algorithm can be improved to better address path heterogeneity, especially for WiFi, LTE, and the emerging 5G Cellular links. We will extend the user-defined routing framework to support finer routing granularity and more flexible forwarding actions. We will also port MPIP to IPv6. Finally, we will study efficiency, fairness and stability of the vertical and horizontal interactions of MPIP with legacy TCP and IP protocols through analysis, simulations and prototype experiments.  

\newpage
\bibliographystyle{IEEEtran}
\bibliography{IEEEfull,mpip_new}
\end{document}

%% file: intro.tex
\vspace{-1.5mm}
\section{Introduction}
\label{sec:intro}
Contemporary end devices are normally equipped with multiple network interfaces, ranging from datacenter blade servers to user laptops and handheld smart devices. Exploiting all available interfaces, applications can adopt multipath transmissions to achieve higher and smoother aggregate throughput, resilience to traffic variations and failures on individual paths, and seamless transition between different networks. While each application can implement its own multipath transmission at the application layer, it is more desirable to provide multipath transmission services from the lower network protocol stack so that all applications can benefit. Recently, Multipath TCP (MPTCP) has been proposed and attracted lots of attention from academia and industry~\cite{mptcp, chen01, chen02, BRBH10, PKB13}. IETF proposed RFC $6182$ specifically for multipath TCP in $2011$. In MPTCP, if a pair of nodes have multiple end-to-end IP paths, each TCP session is carried by multiple subflows, each of which is an independent regular TCP connection on one of the available paths. TCP packets generated by the sender are dispatched to different subflows and transmitted over different paths. At the receiver end, all packets coming from different subflows are put back for reconstructing the original TCP data stream. MPTCP allows all TCP-based applications enjoy the multipath gain in a {\it transparent} fashion. However, UDP-based applications cannot benefit from it. 


{\it In this paper, we share our experience of orchestrating multipath transmission from the network layer of end devices, and present a complete design of Multipath IP Transmission (MPIP).} There are several advantages of implementing multipath transmission at the network layer:

\noindent {\bf Broader Coverage.}  MPIP can transmit IP packets generated by any TCP or UDP based application. Being transparent to the upper layers, MPIP can benefit all user applications without changing the application and transport layer protocols.   

\noindent {\bf  Better View and Coordination.} The network layer can directly measure network status and promptly capture various dynamic events, such as interface and network changes. Since all application traffic go through the network layer, MPIP can efficiently piggyback network measurement on the existing application traffic for {\it in-band measurement}, without generating extra probing traffic. The obtained network information and routing intelligence can be shared cross all applications. MPIP can adjust the transmission strategies for all applications in a coordinated fashion to maximally satisfy the diverse application and user needs. 

\noindent {\bf  More Flexible Routing.} With MPTCP, traffic allocated to a path is determined by the rate achieved by the TCP subflow on that path, i.e., routing is simply determined by congestion control along multiple paths. This is too rigid and limited for applications with diverse throughput and delay needs,  and users with different resource and economic constraints. MPIP instead can implement any customized multipath routing to satisfy application and user needs.  

\noindent {\bf  Lower Complexity.}  MPIP can eliminate redundant network probings and routing adjustments attempted by individual applications and sessions. From the implementation point of view, similar to MPTCP, MPIP only requires changes on end devices. MPTCP has to work with the complexity  resulted from the stateful TCP implementation. The legacy IP protocol is stateless and its implementation is much simpler than the legacy TCP. This leaves more design space for MPIP.  

Meanwhile, MPIP also faces additional challenges. First of all, due to the stateless nature of IP, there is no existing session and path management mechanisms at network layer. Secondly, to work with multiple paths, MPIP constantly needs feedbacks about the availability and performance of each path. However, the legacy IP does not provide end-to-end feedbacks. Thirdly, various middle-boxes, e.g., NAT routers, are {\it by-no-means transparent}. They change and verify IP and TCP headers, and drop packets which they believe are ``unorthodox" according to the legacy TCP/IP protocol. Multipath transmission unavoidably leads to out-of-order packet delivery. This will cause problem for running legacy TCP over MPIP. Finally, MPIP design and implementation should minimize the overhead and complexity added to the network layer. We address those challenges in our MPIP design and implementation. The contribution of our work is three-fold: 

\begin{enumerate*}
\item We develop a complete design to implement multipath transmission at the network layer, consisting of signaling, session and path management, multipath IP source routing, and NAT traversal. Our MPIP design not only can be used by the legacy TCP and UDP protocols, but also works seamlessly with MPTCP.

\item MPIP supports diverse multipath routing strategies. For {\it all-paths} mode, we design a delay-based routing algorithm for MPIP to balance the loads of available paths. We also develop a user-defined multipath routing framework, through which customized routing strategies, such as {\it selected-paths} and {\it single-path}, can be realized by MPIP to satisfy diverse application/user needs.

\item We implement MPIP in Linux and Android kernels. We evaluate its performance using controlled lab experiments and Internet experiments. We demonstrate that MPIP can transparently achieve various multipath gains at the network layer. It works seamlessly with legacy transport layer protocols and popular applications. It can significantly improve user Quality-of-Experience (QoE) using easily configurable multipath routing strategies.
\end{enumerate*}

The rest of the paper is organized as follows. The semantics of MPIP is presented in Section~\ref{sec:sem}. The complete MPIP design is developed in Section \ref{sec:design}. Special issues related to TCP are addressed in Section \ref{sec:tcp}. In Section \ref{sec:evaluation}, we report the experimental results.  Related work is summarized in Section \ref{sec:related}. The paper is concluded in Section \ref{sec:conclusion}.

%% file: design.tex
\vspace{-1.5mm}
\section{Semantics}
\label{sec:sem}
MPIP works at the network layer on end devices. The basic building blocks are: {\it Node, Session, and Path}. 

\begin{itemize}
\item {\it Node} refers to an end device with potentially multiple network interfaces, each of which gets assigned with a private or public IP address. MPIP also works with nodes with single network interface.

\item {\it Session} is a transport layer flow between two nodes served by MPIP. A session is established at the transport layer, using the legacy TCP or UDP protocol, or even the new MPTCP protocol. 

\item {\it Path} is an end-to-end IP route available for a session. For each session, MPIP can use any interface on one node to transmit packets to any interface on the other node. If the two nodes have $m$ and $n$ interfaces respectively, the number of possible paths is $mn$.
\end{itemize}

\begin{figure}
\centering
\includegraphics[width=0.8\linewidth]{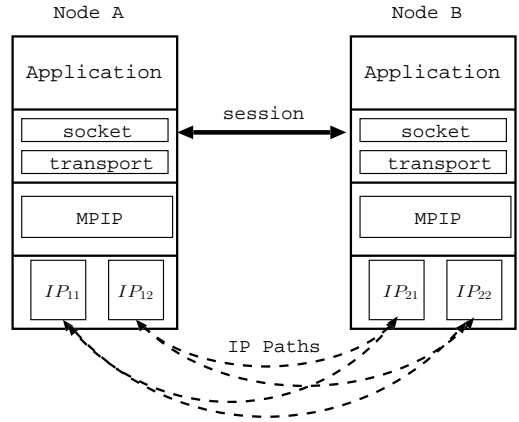}
\caption{Example of MPIP Transmission}
\label{fig.example}
\end{figure}
With the legacy IP, each session is associated with only one IP (interface) and one port number on each node. The routing decision is based on destination IP address. MPIP employs customized {\it session-based} routing, and transmits packets of each session using any combination of the  available paths. For the example in Figure~\ref{fig.example}, node A and node B are MPIP-enabled. They use the legacy application layer and transport layer. Each node has two interfaces (and the associated IP addresses). There are four end-to-end IP paths, as illustrated in Figure~\ref{fig.example}.
When an application on node A opens a TCP/UDP connection to node B, MPIP will treat this connection as a new session. For each packet going from A to B, MPIP will choose one of the four available paths to send it out. To do that, MPIP will change the source and destination IP addresses as well as the port numbers of the packet so that it can be forwarded to the corresponding interface of the chosen path on node B. When node B receives the packet, it will first check which session it belongs to, then modify the IP address and port number back to the original values of the session. Finally, the packet will be passed to the corresponding TCP/UDP socket. The whole process is {transparent} to TCP/UDP session. If MPIP can simultaneously utilize the four paths by dispatching different packets to different paths, TCP /UDP throughput can be improved. Also the session can work normally as long as one path is available, which means TCP/UDP session will not be interrupted even if the default interfaces assigned to the session by the OS are disconnected. This makes hand-overs between different networks seamless and transparent to the transport and application layers. In general, MPIP routes packets from one session using several modes:
\begin{enumerate}
\item {\it all-paths mode:} packets are dispatched concurrently to all the available paths. Each packet will be transmitted along one of the paths. {\it MPIP Routing} determines the traffic splitting ratios among paths. This mode can potentially utilize the bandwidth available on all paths to achieve higher session throughput.  

\item {\it selected-paths mode:} packets are routed on a subset of paths that meet the requirements of the application. Selected-paths mode avoids the inclusion of bad paths that will drag down the application performance. Path selection is application-specific and can be adapted by MPIP based on both application and network dynamics.   

\item {\it single-path mode:} at any time, packets are only routed over one selected path, which can change during the course of the session. MPIP will handle seamless handover between paths, without interrupting the session. Single-path mode eliminates path quality disparity, such as out-of-order packet delivery, by sacrificing the throughput gain, compared with the all-paths and selected-paths modes. 

\item {\it protected-path mode:} a mission-critical packet can be simultaneously transmitted on multiple paths. The receiver will pass the first arrived copy to the upper layer and discard the subsequent redundant copies. It sacrifices bandwidth for resilience. For example, for TCP over two lossy wireless links, ACK packets can be transmitted using the protected-path mode. 
\end{enumerate} 


\vspace{-1.5mm}
\section{MPIP Design}
\label{sec:design}
To realize the gain of MPIP, there are several major design components: {\it Signaling Channel}, {\it Handshake}, {\it Session Management}, {\it Path Management},  {\it MPIP Routing},  and {\it NAT Traversal}. Our design only changes the IP protocol at the network layer and is transparent to the transport and application layers. 
To keep the simplicity of IP protocol, MPIP is still implemented as connectionless, while maintaining some feedback 
information of the available paths necessary for MPIP routing. We achieve this by simply keeping track of several key 
tables. 

\subsection{Signaling Channel}
 \begin{table}[htbp]
\caption{\label{tb.cm}Control Message Block}
\centering
{ \begin{tabular}{|c|c|c|c|}
\hline
\rowcolor{LightCyan}
\emph{Source} & \emph{Session} & \emph{Local IP}& \emph{CM}\\
\rowcolor{LightCyan}
\emph{Node ID} & \emph{ID} & \emph{Address List}& \emph{Flags} \\
\hline
\rowcolor{LightCyan}
\emph{Path} & \emph{Feedback} & \emph{Packet} & \emph{Path}\\
\rowcolor{LightCyan}
\emph{ID} & \emph{Path ID} & \emph{Timestamp} & \emph{Delay}\\
\hline
\end{tabular}
}
\end{table}
In TCP protocol, ACK packets are used to feedback information from the receiver. Due to its connectionless design, 
IP protocol doesn't have its built-in end-to-end feedback channel. MPIP routing algorithms do need realtime information 
about the availability and performance of end-to-end paths. We need a signaling channel for MPIP. Instead of 
transmitting extra signaling packets, we piggyback MPIP control information to each MPIP packet.

For each packet sent out by MPIP, we add an additional control message (CM) data block at the end of user data.
The size of the CM block is $25$ bytes, a small overhead for typical data packets of $1000+$ bytes. 
Considering the throughput gain and robustness brought by MPIP, the overhead of CM block is well acceptable. Packet size may exceed the link MTU after attaching the CM block. We force the transport layer to reduce the size of each segment, e.g. decreasing the MSS value for TCP connection, to make sure the CM block fits within the MTU limit.
The information contained in a CM block of a packet is shown in Table~\ref{tb.cm}.

\emph{Source Node ID} is a globally unique ID of the sending node of this packet. Since each node 
has multiple interfaces, and their IP addresses may change over time, we should not use interface IP addresses 
to identify a node. To have a semi-static node ID, we instead use the MAC address of a NIC (preferable more 
static ones) on the node to be its ID. 

\emph{Local IP Address List} carries all local IP addresses on the sending node. This list will be used to 
construct MPIP paths.

\emph{CM Flags} encodes the MPIP functionality of the packet. With different values of \emph{CM Flags}, 
different actions will be operated when the packet is received.

Other fields will be explained in the following sections.
\subsection{Handshake}
\label{sec:handshake}
\begin{table}[htbp]
\caption{\label{tb.me}MPIP Availability}
\centering
\begin{tabular}{|c|c|c|c|}
\hline
\rowcolor{LightCyan}
Dest. IP 		& Dest. Port 	 & MPIP  		   & Query\\
\rowcolor{LightCyan}
Address & Number & Availability & Count\\
\hline
${IP}_{1}$ & ${P}_{1}$ & \texttt{True}  & $2$ \\
\hline
${IP}_{2}$ & ${P}_{2}$ & \texttt{False} & $5$ \\
\hline
\end{tabular}
\end{table}

\begin{figure}
\centering
\includegraphics[width=0.8\linewidth]{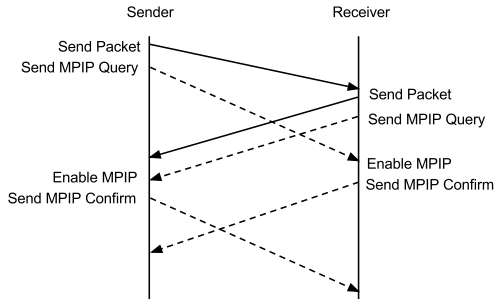}
\caption{MPIP Handshake}
\label{fig.handshake}
\end{figure}
As an extension of IP, MPIP needs to be backward compatible. To take advantage of MPIP, both end nodes 
of a session need to be MPIP enabled. Locally, every node maintains a table (Table~\ref{tb.me}) to record 
the availability of MPIP on remote nodes. Due to the existence of NAT, two nodes may share 
the same IP address, this is why we have to index each entry using the combination of IP address and port number.
The MPIP handshake process is illustrated in Figure~\ref{fig.handshake}.
When a node receives a packet from the transport layer, it first checks locally whether the destination address 
and port number has an entry in Table~\ref{tb.me}. If {\it yes} and MPIP availability is \texttt{True} , then the packet will be 
sent out using MPIP; if MPIP availability is \texttt{False}, it will be sent out as a normal IP packet to be backward 
compatible. If there is no entry found in the table, besides sending out the packet as a normal IP packet, MPIP 
makes a copy of the packet and inserts the CM block with $Flags\_Enable$. This value is used for MPIP query. 
When the packet is received by a MPIP-enabled node, the receiver adds the sender's IP address and port number 
into its own MPIP availability table with value of \texttt{True}, then sends back a confirmation packet to the sender 
with $Flags\_Enabled$. When the sender receives the confirmation, it will add the receiver's address and port 
number to its local MPIP availability table.
For TCP connections, the confirmation packet generated by MPIP receiver maybe blocked by NAT device if its sequence number 
is not properly set. We solve this problem by piggybacking it on a duplicated regular TCP packet. Please refer to our technical report 
for more detail~\cite{MPIP_Tech}.  
%

In Table~\ref{tb.me}, the column \emph{Query Count} maintains the number of query messages that have been sent out to each destination. 
If the number is larger than a threshold value, it assumes that the destination doesn't support MPIP, and marks 
the availability in the table as \texttt{False}. 
%

\begin{table}[htbp]
\caption{\label{tb.wi}Node ID vs IP address and Port\label{tb.id}}
\centering
\begin{tabular}{|c|c|c|}
\hline
\rowcolor{LightCyan}
 Node ID  & IP Address & Port Number\\
\hline
${ID}_{1}$&${IP}_{11}$&${P}_{11}$ \\
\hline
\rowcolor{Gray}
${ID}_{1}$&${IP}_{12}$&${P}_{12}$ \\
\hline
${ID}_{2}$&${IP}_{21}$&${P}_{21}$ \\
\hline
\rowcolor{Gray}
${ID}_{2}$&${IP}_{22}$&${P}_{22}$ \\
\hline
\end{tabular}
\end{table}

After the MPIP handshake, a node can start to learn the interfaces available on each MPIP-enabled remote node.
 Each node maintains a node ID to IP address and port number mapping table (Table~\ref{tb.wi}). Every time a 
 MPIP packet is received, the receiver extracts the sender's node ID from the  packet's CM block, 
 and IP address and port number from the packet header. The three tuple is then written into the mapping table. 
 


%
%
%

\begin{table*}
\caption{\label{tb.ss}Session Information Table}
\centering
\begin{tabular}{|c|c|c|c|c|c|c|c|c|c|}
\hline
\rowcolor{LightCyan}
Dest.  & Session &  Source &  Source & Destination & Destination & Protocol  &    Next      & Update     \\
 \rowcolor{LightCyan}
 Node ID  &   ID    &    IP   &   Port  &     IP      &    Port     &           &  Sequence No &  Time    \\
\hline
${ID}_1$&${SID}_1$&${SIP}_{1}$&${SPORT}_{1}$&${DIP}_{1}$&${DPORT}_{1}$&TCP&$S_1$&$T_1$               \\
\hline
\rowcolor{Gray}
${ID}_1$&${SID}_2$&${SIP}_{1}$&${SPORT}_{2}$&${DIP}_{1}$&${DPORT}_{2}$&UDP&$0$&$T_2$                 \\
\hline
${ID}_2$&${SID}_1$&${SIP}_{2}$&${SPORT}_{3}$&${DIP}_{2}$&${DPORT}_{3}$&TCP&$S_2$&$T_3$              \\
\hline
\rowcolor{Gray}
${ID}_2$&${SID}_2$&${SIP}_{2}$&${SPORT}_{4}$&${DIP}_{2}$&${DPORT}_{4}$&UDP&$0$&$T_4$                 \\
\hline
\end{tabular}
\end{table*}

\subsection{Session Management}
\label{sec:session}
MPIP conducts session-based routing. Session management takes care of the addition and removal of TCP and UDP sessions.
At the transport layer, each session is identified by the traditional $5$-tuple: source and destination 
IP addresses and port numbers, and protocol type.  Since MPIP can transmit a packet from a session using 
different source and destination IP address/port numbers than the session's original ones, we can no longer 
use IP addresses/port numbers to associate a MPIP packet with a transport layer session. Instead, we will use session 
ID and node ID carried in the CM block to identify the session of a MPIP packet. We need a table to correlate 
the two different session mapping schemes employed by MPIP and the legacy transport layer. This is achieved 
through the session information table, as in Table~\ref{tb.ss}.

The table maintains one entry for each session to each remote node. For each entry, the socket information, 
namely IP addresses and port numbers, are the original ones from the transport layer. A session's socket 
information will not be changed even if the IP addresses and port numbers that are initially assigned to the 
session are no longer active. This is for seamless hand-overs between networks.

After the MPIP availability handshake has been successfully completed, when sending out a packet, the sender 
checks Table~\ref{tb.ss} to see whether a proper session entry has been generated. If not, MPIP generates a 
new session ID and adds a new entry to Table~\ref{tb.ss}. The IP addresses, port numbers and protocol are 
extracted from the packet header, and the destination node ID is obtained from Table~\ref{tb.wi}.
After this, all packets belong to the session will carry the session's ID in its CM block.
On the receiver end, whenever a MPIP packet is received, the receiver extracts the source node ID and session 
ID from its CM block. If there is no entry found in its session information table, it will generate a new entry 
and populate it with the source node ID, session ID, and socket information carried in the packet header, with 
swapped source and destination IP/port addresses. This will make sure that both sides of the same session use 
the same session id. Note that, due to NAT, for the same session, the IP addresses and port numbers seen by a 
remote node might be different from the values on a local node. This won't cause any confusion as long as the 
session ID and node ID combination is unique.

Removal of a session is done by expiration based on the session's \emph{Update Time} in Table~\ref{tb.ss}.  
The column \emph{Next Sequence No} is used for TCP out-of-order process which will be explained in Section~\ref{sec:outoforder}.

\subsection{Path Management}
\label{sec:path}

\begin{table*}
\small
\caption{\label{tb.pi}Path Information Table}
\centering
\begin{tabular}{|c|c|c|c|c|c|c|c|c|c|c|c|}
\hline
\rowcolor{LightCyan}
 Dest & Session &  Path  & Src &   Src & Dest & Dest &   Minimum       & Real-Time      & Real-Time      & Maximum       & Path   \\
\rowcolor{LightCyan}
Node ID  &   ID  & ID & IP  &  Port &  IP  & Port &  Path Delay  & Path Delay  & Queuing Delay  & Queuing Delay & Weight \\
\hline
$ID$&${SID}_{1}$&${PID}_{11}$&${sip}_{1}$&${sp}_{1}$&${dip}_{1}$&${dp}_{1}$&${D_{min}}_{11}$&$D_{11}$&${Q}_{11}$&${{Q}_{max}}_{11}$&$W_{11}$\\
\hline
\rowcolor{Gray}
$ID$&${SID}_{1}$&${PID}_{12}$&${sip}_{2}$&${sp}_{2}$&${dip}_{1}$&${dp}_{1}$&${D_{min}}_{12}$&$D_{12}$&${Q}_{12}$&${{Q}_{max}}_{12}$&$W_{12}$\\
\hline
$ID$&${SID}_{2}$&${PID}_{21}$&${sip}_{1}$&${sp}_{1}$&${dip}_{1}$&${dp}_{1}$&${D_{min}}_{21}$&$D_{21}$&${Q}_{21}$&${{Q}_{max}}_{21}$&$W_{21}$\\
\hline
\rowcolor{Gray}
$ID$&${SID}_{2}$&${PID}_{22}$&${sip}_{2}$&${sp}_{2}$&${dip}_{2}$&${dp}_{2}$&${D_{min}}_{22}$&$D_{22}$&${Q}_{22}$&${{Q}_{max}}_{22}$&$W_{22}$\\
\hline
\end{tabular}
\end{table*}
After a session is registered with MPIP, the next step is to explore all the available paths for the session. One simple solution is to have each node send their local IP addresses to the other end using the {\it Local Address List} in CM block. Then any pair of IP addresses on the two ends can be used as a path for MPIP transmission. However, this only works if all interfaces on both ends have public IP addresses. If one node is behind a NAT, its local IP addresses cannot be
used directly to establish IP paths. To solve this problem, we again have to identify paths using a combination of IP address and port number on both ends. Consequently, the path management has to be done for each session individually.

\subsubsection{Establishment}
\label{sec:estab}
MPIP maintains a path information table on each node, as in Table~\ref{tb.pi}, to record the available paths for each session. Each entry contains the ID of the remote node and the session ID. Each path is allocated with a path ID, which is unique on the local node. 
The source and destination IP and port addresses are the addresses carried in the
header of MPIP packet, NOT necessarily the same as those allocated to the session at the transport layer.

\begin{figure}
\centering
\includegraphics[width=\linewidth]{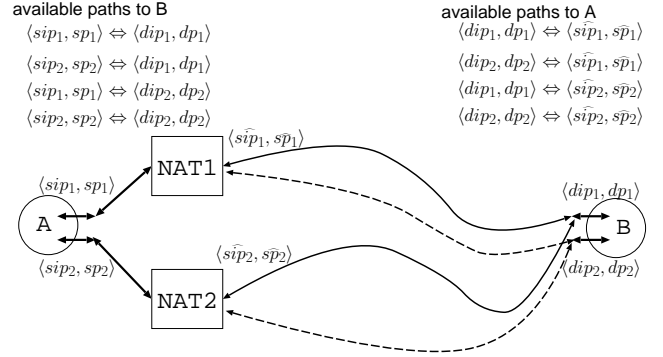}
\caption{MPIP Path Establishment with NAT}
\label{fig.path}
\end{figure}

Given $m$ and $n$ interfaces at each end node, there are totally $mn$ possible paths. After the MPIP handshake, each node tries to send out packets from each of its local interface to each of the known interface on the remote node. If a packet with a certain combination of source and destination IP/port addresses can get through, the node will add the path to path information table. Let's explain the process through the example in Figure~\ref{fig.path}. Node A initiates a session with node B. The IP and port addresses allocated to the session at the transport layer are $\langle sip_1, sp_1 \rangle$ and $\langle dip_1, dp_1 \rangle$ on A and B respectively. Without loss of generality, let's assume the session can be established correctly with legacy IP. Then on both ends, MPIP records the new session, and adds the default path between $\langle sip_1, sp_1 \rangle$ and $\langle dip_1, dp_1 \rangle$ for the session in Table~\ref{tb.pi}. Since A knows B is MPIP-enabled, it also tries to send the same packet from its other local interface with IP address $sip_2$ by changing its source addresses to $\langle sip_2, sp_2 \rangle$. When B receives the packet, possibly due to NAT, the source IP and port addresses in the packet might be different from $\langle sip_2, sp_2 \rangle$, say $\langle \widehat{sip_2},\widehat{sp_2} \rangle$. Then B examines the {\it Source Node ID} and {\it Session ID} in the packet's CM block, it knows this is a MPIP transmission for the same session but from a different interface. B adds for the session a new path  with destination address of $\langle \widehat{sip_2},\widehat{sp_2} \rangle$ in its path information table. Now B will also send back packets to A's second interface, using destination addresses $\langle \widehat{sip_2},\widehat{sp_2} \rangle$.
%
When A receives the packet, it confirms the connectivity of its local path between $\langle sip_2, sp_2 \rangle$ and $\langle dip_1, dp_1 \rangle$, and adds it to its path information table. Similarly, if B has another interface with public address $dip_2$, A will obtain the new address from the {\it Local Address List} in the CM block of packets from B to A. Then A can establish more IP paths to this new address using a similar process.

%
\subsubsection{Monitoring}
\label{sec:feedback}
To facilitate path selection, MPIP continuously monitors the performance of active paths. Given that packet losses in the current Internet are rare, we mainly focus on path delay in our current design. Due to asymmetric routing and unequal congestion levels along two directions of the same path, instead of measuring the round-trip delay of a path, we measure the one-way path delay to infer the path quality on each direction.
%
In Table~\ref{tb.pi}, all fields related to network delay will be calculated by one-way path delay feedback from the remote node. When node $A$ sends out a packet, it chooses a path from Table~\ref{tb.pi} and sets \emph{Packet Timestamp} with its local system time $T_1$. After node $B$ receives this packet, it calculates the one-way delay for the path from $A$ to $B$ as $T_2-T_1$, where $T_2$ is B's local time when receiving the packet. 
In practice, the absolute value of path delay calculated here isn't the real delay value because of the clock difference between node $A$ and node $B$. But our path selection algorithms depend on the relative ordering and variations of path delays, instead of their absolute values. Clock difference between nodes has little impact. $B$ then sends back the path delay information in the CM block of the next packet going back to $A$, which records the path delay value into the column \emph{Real-Time Path Delay} in Table~\ref{tb.pi}. Path delay values are smoothed using a simple moving average algorithm. More details can be found in our technical report~\cite{MPIP_Tech}


\subsubsection{Dynamic Path Addition and Removal}
\label{sec:switch}

As multipath feature enabled on a device, IP addresses of interfaces change dynamically. A mobile device can connect to different access points (WiFi hotspot/Cellular Tower) during a session. Its IP addresses can be changed, removed or added back dynamically. To make the changes transparent to applications, MPIP supports dynamic addition and removal of paths from Table~\ref{tb.pi}.
When IP address change happens on one node, it sets $Flags\_IP\_Change$ in the CM block of its next outgoing packet. After receiving a packet with this flag, the receiver knows that IP address on the sender has changed, it removes all path entries  
related to the changed IP address in Table~\ref{tb.pi}. Meanwhile, the entry for this session in Table~\ref{tb.ss} remains unchanged. The path that sends out the IP change notification will be added back to the aforementioned tables as the only path of the session. Also, the sender does the same reset for this session. After all these resets, there is only one path left for this session, all the other available paths will be added back through the procedure in Section~\ref{sec:estab}. Similarly, when a new interface becomes available, new IP paths from it can be added using the the mechanism in Section~\ref{sec:estab}. 


%

\subsubsection{Periodical Heartbeat}
Table~\ref{tb.pi} should be updated continuously on both sides. During the lifetime of a TCP session, both sides send packets to each other at a high frequency. However UDP doesn't have this built-in feedback mechanism and in some UDP applications, all traffic is one-way without acknowledgement, which means that the sender can't get feedback information through piggyback. 
To solve this problem, a periodical heartbeat mechanism is introduced to keep Table~\ref{tb.pi} fresh. More details can be found in our technical report~\cite{MPIP_Tech}.  

\subsection{Multipath IP Source Routing}
\label{sec:selection}
Given all paths available for a session, every time one node needs to send out a packet, it chooses the most suitable path from Table~\ref{tb.pi}. MPIP offers different routing strategies to satisfy the diverse needs of applications.

\subsubsection{All-paths Mode}
\label{sec:delay}
Many applications, e.g., web, file transfer, and video streaming, can benefit from high-throughput transmissions. MPIP can concurrently transmit packets along multiple paths to achieve higher throughput than the traditional single path routing. With MPTCP, the transmission rate along each TCP sub-flow is controlled by the TCP congestion control algorithms. Since MPIP works under rate control schemes from transport and application layers, it will be redundant and possibly conflicting to implement fine-grained rate control for each MPIP path at the network layer. Instead, the main design goal of MPIP routing is to balance load along concurrent paths using end-to-end path delay feedback and probabilistic packet dispatching algorithm.

As in Table~\ref{tb.pi}, we maintain a  \emph{Path Weight (W)} for each active path. Each packet will be dispatched to a path $k$ with the probability $P(k)$, which is calculated as:
\be
\label{eq.choose}
P(k) = \frac{W_k}{\sum_{i=1}^{N}W_i}.
\ee
Path weight is the only criterion for path selection and determines the performance of MPIP load balancing. 
In our prototype, we use realtime one-way path delay to dynamically update path weights.


End-to-end path delay consists of propagation delay, transmission delay, processing and queueing delay. While propagation 
delay and transmission delay are mostly constant, processing and queue delay are time-varying and increase with congestion 
level. 
We maintain the minimum path delay to represent the constant portion of end-to-end path delay, and use the difference between real-time and minimum delay to infer the queuing delay, which reflects the congestion level along the path. We then adjust the weight of each path using the 
real-time queuing delay.

In Table~\ref{tb.pi}, \emph{Real-Time Path Delay} $D$ is collected using receiver feedbacks as described in 
Section~\ref{sec:feedback}. Every time a new path delay sample $D$ is received, the other three delay metrics are updated 
as follows.
\begin{enumerate*}
\item {\it Minimum Path Delay:} $D_{min}=\min{\{D_{min}, D\}}$;
\item {\it Real-Time Queuing Delay:} $Q=D-D_{min}$;
\item {\it Maximum Queuing Delay:}  $Q_{max}=\max{\{Q_{max}, Q\}}$.
 \end{enumerate*}

During our experiments, we found that calculating the weight of each path independently according to its 
realtime queuing delay can result in large fluctuations. We instead adjust the weights of all paths together 
based on their queueing delay variations as in Algorithm~\ref{alg.bw}.
\begin{algorithm}
\caption{Path Weight Adjustment.}
\label{alg.bw}
\begin{algorithmic}[1]
\STATE  $Q_{avg}= \frac{\sum_{i=1}^{N}Q_i}{N}$; //{\it average delay among all paths}
\IF {$Q_i\le Q_{avg}$}    \STATE $W_i=W_i+S$; //{\it increase weight for low delay path}
    \IF {$W_i>1000$}
	    \STATE $W_i=1000$; //{\it upper bound for path weight} 
	\ENDIF
\ELSE
    \STATE $W_i=W_i-S$; //{\it decrease weight for high delay path}
    \IF {$W_i<1$}
	    \STATE $W_i=1$; //{\it lower bound for path weight} 
	\ENDIF
\ENDIF
\RETURN;
\end{algorithmic}
\end{algorithm}
$N$ is the number of paths that belong to one session, $Q_i$ and $W_i$ are queuing delay and weight of path $i$, and $S$ is the adjustment granularity. Initially, every path has the same path weight of $\frac{1000}{N}$. In each iteration, the path weight increases or decreases by $S$ based on whether its queuing delay is higher or lower than the average delay. The maximum weight is $1000$, and the minimum is $1$. This way, we keep all live paths in consideration. Heavily congested paths will not be completely eliminated. Instead they will have the minimum weight, and their weights will be increased after congestion is relieved. Algorithm~\ref{alg.bw} is executed periodically, the length of each period is defined as a configurable system parameter $T$. Now we have two configurable parameters: the adjustment stepsize $S$ and interval $T$. Larger $S$ and shorter $T$ react faster to congestion level changes, but generate larger fluctuations; while smaller $S$ and longer $T$ can result in smaller fluctuation but sluggish response. In our system, for the path weight range of $1 \sim 1000$, we set $S$ to $10$ and $T$ to $100$ ms. During our evaluation, this configuration achieves good balance between fluctuation and convergence.

\subsubsection{User-defined Multipath Routing}
\label{sec:resp}
Not all applications take throughput as the first priority. For a live video streaming session, as long as the throughput is higher than the video rate,
delay is more critical for the streaming quality. Even for the same application, different data may have different QoS requirements.
In the example of video calls, such as WebRTC, audio stream has low volume but are very sensitive to delay, video stream has
high volume and can be less sensitive to delay than audio. To address the diverse needs of applications, we design MPIP to support user-defined
routing schemes, including {\it selected-paths}, {\it single-path} and {\it protected-path}. 
%
%
%
%
%
%
Users can inform MPIP of their desired multi-path routing policies by configuring a routing table as illustrated in Table~\ref{tb.route}.
\begin{table}[htbp]
\caption{\label{tb.route} User-defined Multipath Routing Table}
\centering
\begin{tabular}{|c|c|c|c|c|c|}
\hline
\rowcolor{LightCyan}
 IP   		&   Port   &          & Start  & End    & Routing       \\
 \rowcolor{LightCyan}
 Address    &  Number     & Protocol & Size   & Size & Priority       \\
\hline
$*$&$22$&$TCP$&$0$&$200$&$R_f$   \\
\hline
\rowcolor{Gray}
$192.168.1.2$&$5222$&$UDP$&$200$&$*$&$T_f$   \\
\hline
$192.168.1.2$&$5221$&$UDP$&$0$&$500$&$R_f$  \\
\hline
\end{tabular}
\end{table}
Each line of the table is a customized routing rule for outgoing packets. Each rule matches a set of packets and the routing priority for the matched packets. Packet matching is done using destination IP address, port number, protocol, and the range of packet length. We currently define two types of routing priorities: {\it throughput-first} $T_f$, and {\it responsiveness first} $R_f$. Outgoing packets with $T_f$ priority will be dispatched to available paths using the {\it all-paths} mode presented in Section~\ref{sec:delay}. Outgoing packets with $R_f$ priority will always be sent to path with the lowest delay using the {\it single-path} mode. For example, based on the first row of Table~\ref{tb.route}, for any TCP connection with destination port $22$ (ssh session), if the packet length is smaller than $200$ bytes, the packet will be forward to the lowest delay path. The second row defines that all UDP packets going to a remote host with packet size larger than $200$ bytes should be forwarded using {\it all-paths} mode. The third row specifies that for a UDP packet going to the same remote host, but a different port number, if the packet size is less than $500$, it will be forwarded to the lowest delay path instead.
%
%
%
The current implementation employs rigid packet matching rules and has limited routing policies. Under the same basic framework, we will extend it to  incorporate more flexible and more user-friendly packet matching rules and more diverse routing policies with finer granularity in our future work.
\section{TCP-related Issues}
\label{sec:tcp}
By deviating from the default single-path transmission, MPIP also brings some new issues for the upper layer protocols, especially TCP,  such as NAT checking and out-of-order packet delivery. It is also intriguing to explore the co-existence of MPIP with multi-path transmissions at upper layers, such as MPTCP. We now present solutions to TCP-related issues.

\subsection{NAT Checking}
Based on our experiments and other studies, e.g.~\cite{mptcp}, NAT devices are by no means transparent, and conduct all kinds of mapping, verification, and dropping to end-to-end sessions, especially TCP. One immediate obstacle introduced by NAT to MPIP is that many NAT devices drop a TCP packet if they don't have a record about the TCP connection that the packet belongs to. This doesn't cause a problem for MPTCP since each sub-flow in MPTCP is a legitimate TCP connection, and all packets of a sub-flow, including the three-way handshake packets establishing the sub-flow, traverse the same NAT. In MPIP, if we transmit TCP packets on a path different from the original one through which the TCP connection is established, NAT devices along the path are not aware of the connection and will drop these packets before they arrive at the destination. We provide two solutions.

\subsubsection{Fake TCP Handshake}
To work around a NAT device that drops packets of a TCP connection established on a different path, we construct a fake TCP three-way hand-shake on  the NAT's path before sending packets over. All handshake packets have \emph{CM Flags} set to $Flags\_HS$. They are dropped after being processed by MPIP. As shown in Table~\ref{tb.cm}, the field \emph{Local Address List} carries all local IP addresses. Also, the node that initiates the connection is considered as the client. When the client receives the IP address list of the server, it sends out a SYN packet along each possible path to the server except the original one which was used to initiate the real TCP connection. When the server receives a SYN packet, it replies with a SYN-ACK packet. After the client sends out the final ACK packet to the server, the three-way handshake is completed successfully. After this, NAT routers along the path have a record about this fake TCP connection, will pass TCP packets assigned to the path.

\subsubsection{UDP Wrapper}
The other solution is UDP wrapper. During our experiments, most NAT devices don't verify socket information of UDP packets. We make use of this feature and transmit a TCP packet inside a UDP packet to pass NAT checking. At the sender side, every time the network layer gets a TCP packet from transport layer, MPIP chooses a path to send the packet out as shown in Section~\ref{sec:path}. If the chosen path isn't the original path, we encapsulate the whole TCP packet into an UDP packet by adding a forged UDP header using the corresponding IP addresses and port numbers of the chosen path. At the receiver end, MPIP can tell this UDP packet is a carrier for a TCP packet instead of a regular UDP packet by checking the \emph{Protocol} field of the path in Table~\ref{tb.ss}. After removing the UDP header, the original socket information will be extracted from Table~\ref{tb.ss} to be filled into the TCP and IP headers.


\subsection{Out-of-order Packet Processing}
\label{sec:outoforder}
Different interfaces take different network accesses and different Internet paths to reach the same destination.  Packets sent over multiple interfaces/paths can arrive at the destination node out of order. This is not a problem for protocols like UDP, but for TCP, out-of-order packet delivery will significantly degrade its performance. 
When TCP works over MPIP, if the delay difference between multiple paths is significant, we can expect a lot of out-of-order packets.
To resolve this problem, for each session in Table~\ref{tb.ss}, if it is TCP protocol, MPIP maintains the sequence number $S$ of the next in-order packet of the session to be received. MPIP also maintains a separate buffer $B$ for each active session to store out-of-order packets.
Whenever a new packet is received, if the sequence number is larger than $S$, it will be stored in $B$; if the sequence number equals to $S$, MPIP checks how many consecutive packets are stored in $B$, starting from sequence number $S$. Then MPIP pushes all consecutive packets to the transport layer and update $S$ accordingly.
If one packet is lost or delayed for a long time, all subsequent packets will get stuck in the buffer. As a result, TCP layer will assume that all packets are lost, this will result in catastrophe. To avoid this, we limit the buffer size. All the packets in the buffer will be pushed up once the buffer is full. In our prototype, we set the maximum buffer size to $100$ packets.
\subsection{MPTCP over MPIP}
\label{sec:together}
MPTCP exploits the multi-path gain at the transport layer. A MPTCP session employs multiple subflows, each of which is a legitimate TCP connection over a single IP path. When MPTCP runs over MPIP, each TCP subflow can now utilize multiple paths. For the example in Figure~\ref{fig.example}, a MPTCP session can have $4$ subflows. MPIP will treat each subflow as an independent TCP session, and will create $4$ paths for  each subflow. As a result, there are totally $4$ sessions and $16$ paths managed by MPIP. Now MPTCP and MPIP work together to adapt the traffic allocated to each path. When congestion accumulates on one path, MPIP will first notice the high queuing delay on that path,  reduce the path weight and shift packets to less congested paths. The load balancing conducted by MPIP at the network layer makes the congestion variations along different paths less perceivable for MPTCP subflows so that MPTCP  can make better use of subflows to achieve higher throughput.
We will demonstrate this using MPTCP+MPIP experiments in Section~\ref{sec:tcptp}.

%% file: evaluation.tex
\vspace{-1.5mm}
\section{Performance Evaluation}
\label{sec:evaluation}
To evaluate the performance of the proposed design, we implement MPIP in Linux kernel $3.10.11$ in Ubuntu system. The prototype is designed for IPv$4$. The main functionality is implemented in three new files with more than $5,000$ lines of code. We modified ``ip\_input.c'' and ``ip\_output.c'' under IPv$4$ folder to embed MPIP features into the existing TCP/IP stack. MPIP is also implemented into Android system $6.0.1$ with kernel version $3.10.73$. For all TCP experiments, we use CUBIC-TCP \cite{cubic}. MPTCP version $0.92$ is used in our evaluation. We use {\it Iperf/Iperf3} to generate traffic.

\subsection{Controlled Lab Experiments}
\label{sec:lab}
In our lab, we install the prototype on two desktop computers, which are connected directly to a router. Each desktop has two $100$Mbps NICs, leading to $4$ paths with aggregate capacity of $200$Mbps. We use {\it tc (traffic control)} tool in Linux to control bandwidth and delay on each path.

\subsubsection{TCP over MPIP}
\label{sec:tcptp}
\begin{figure*}[htb]
\centering{
\subfigure[Load Balancing 
\label{fig.twopathstpcomp}]{\includegraphics[width=0.23\linewidth,height=1.4in]{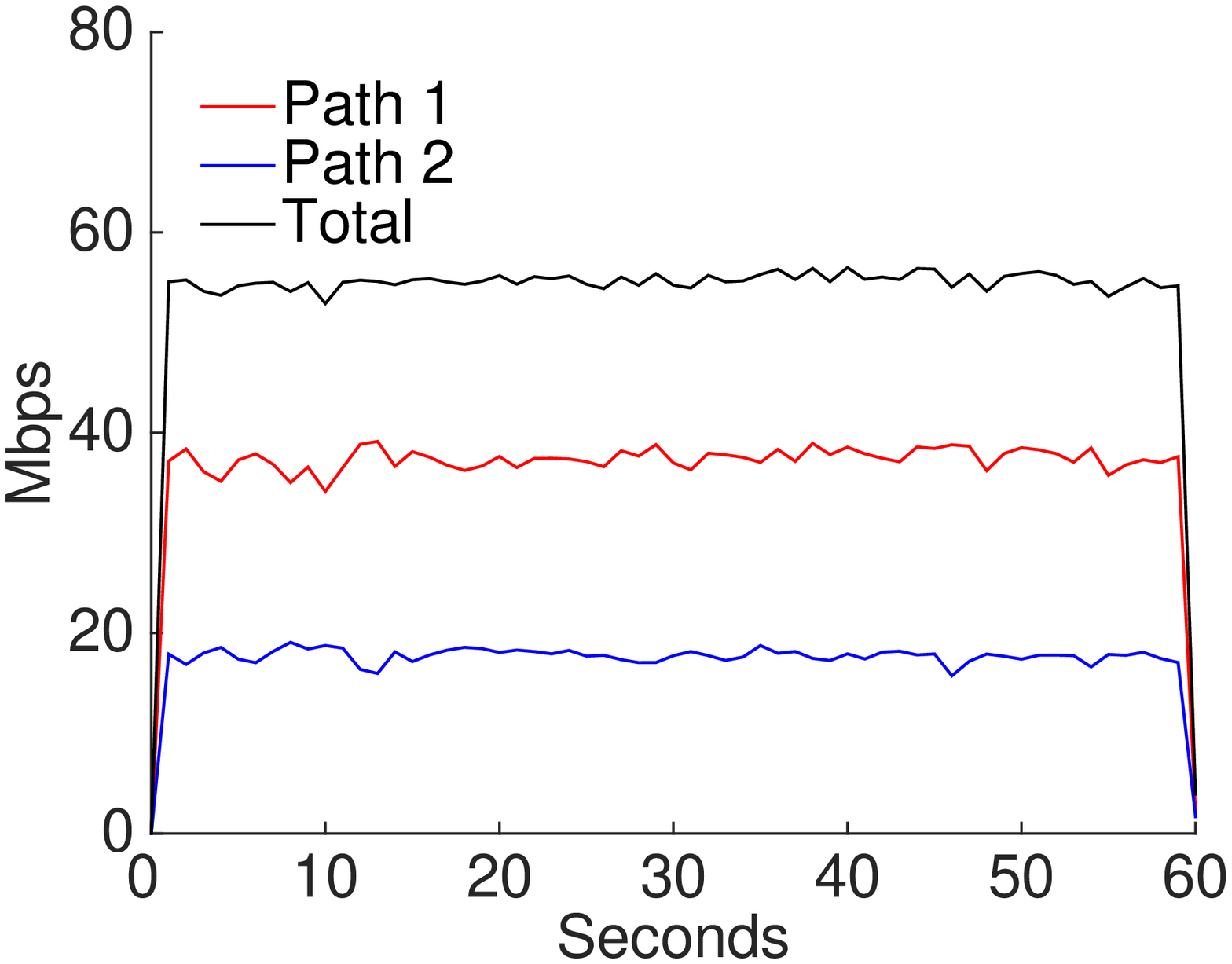}}
\subfigure[Path Throughput under failure
\label{fig:disconnection}]{\includegraphics[width=0.23\linewidth,height=1.4in]{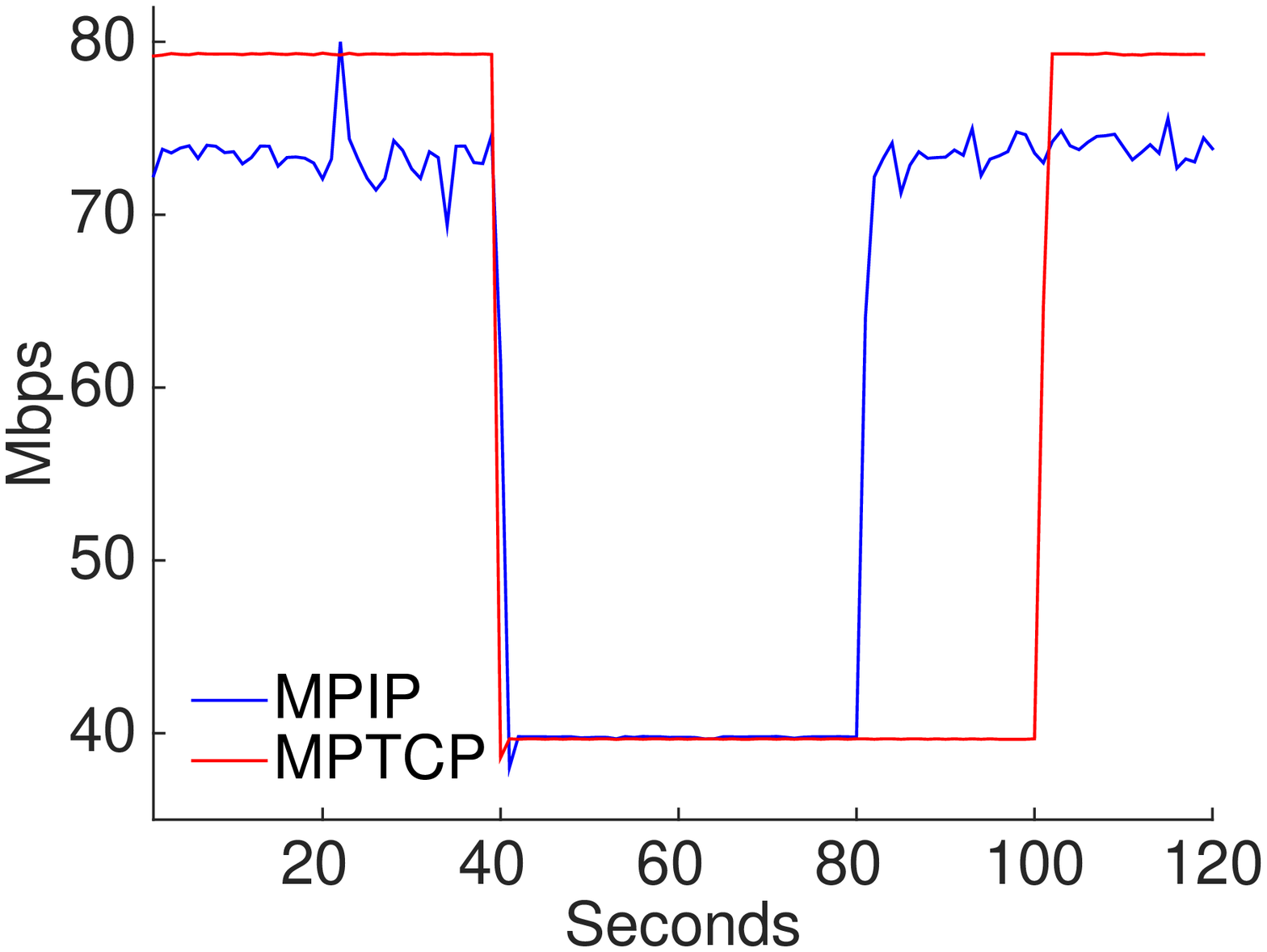}}
}
\subfigure[MPIP Reordering \label{fig:mpip_reorder}]{\includegraphics[width=0.23\linewidth,height=1.4in]{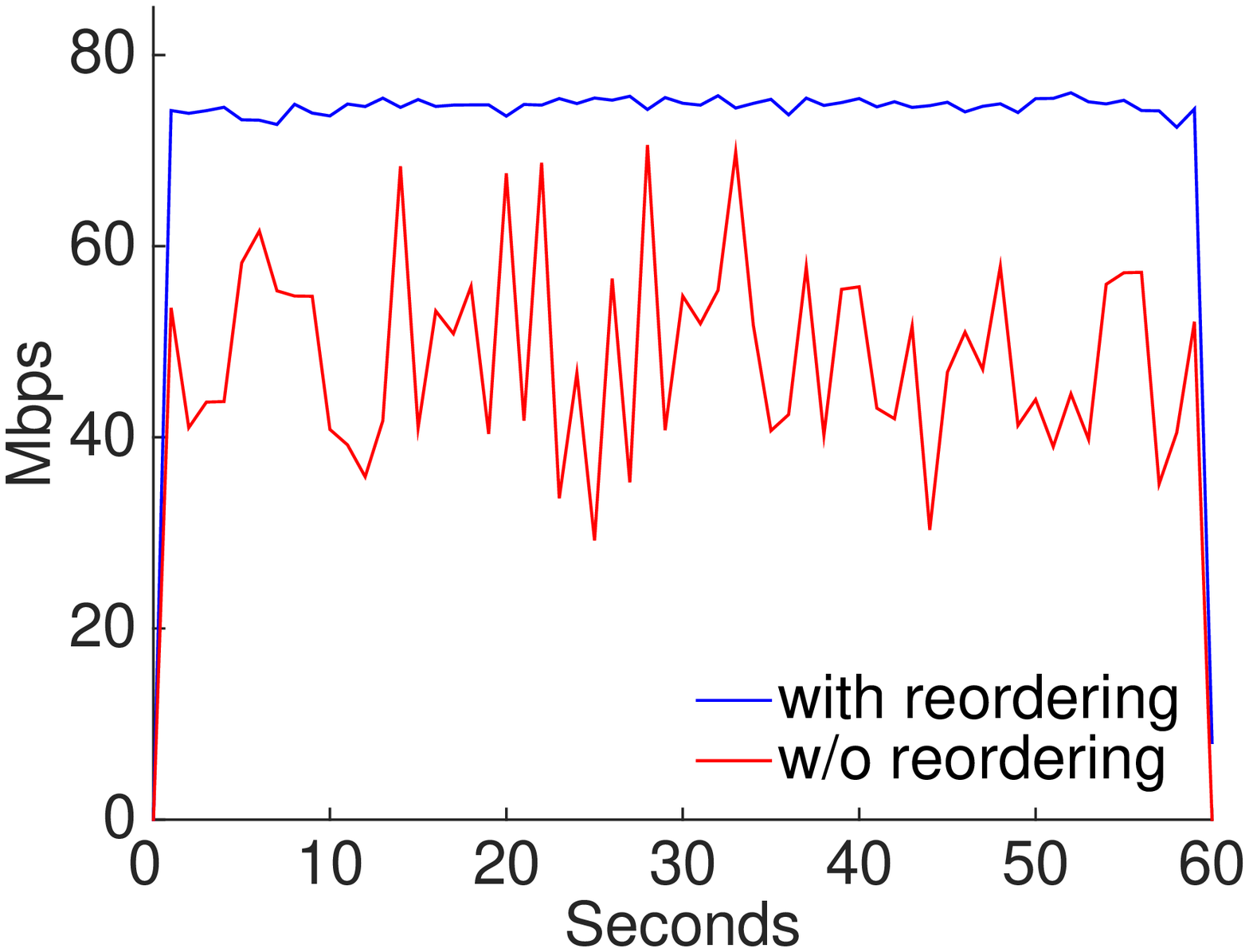}}
\subfigure[MPIP and MPTCP \label{fig:mpip_mptcp}]{\includegraphics[width=0.23\linewidth,height=1.4in]{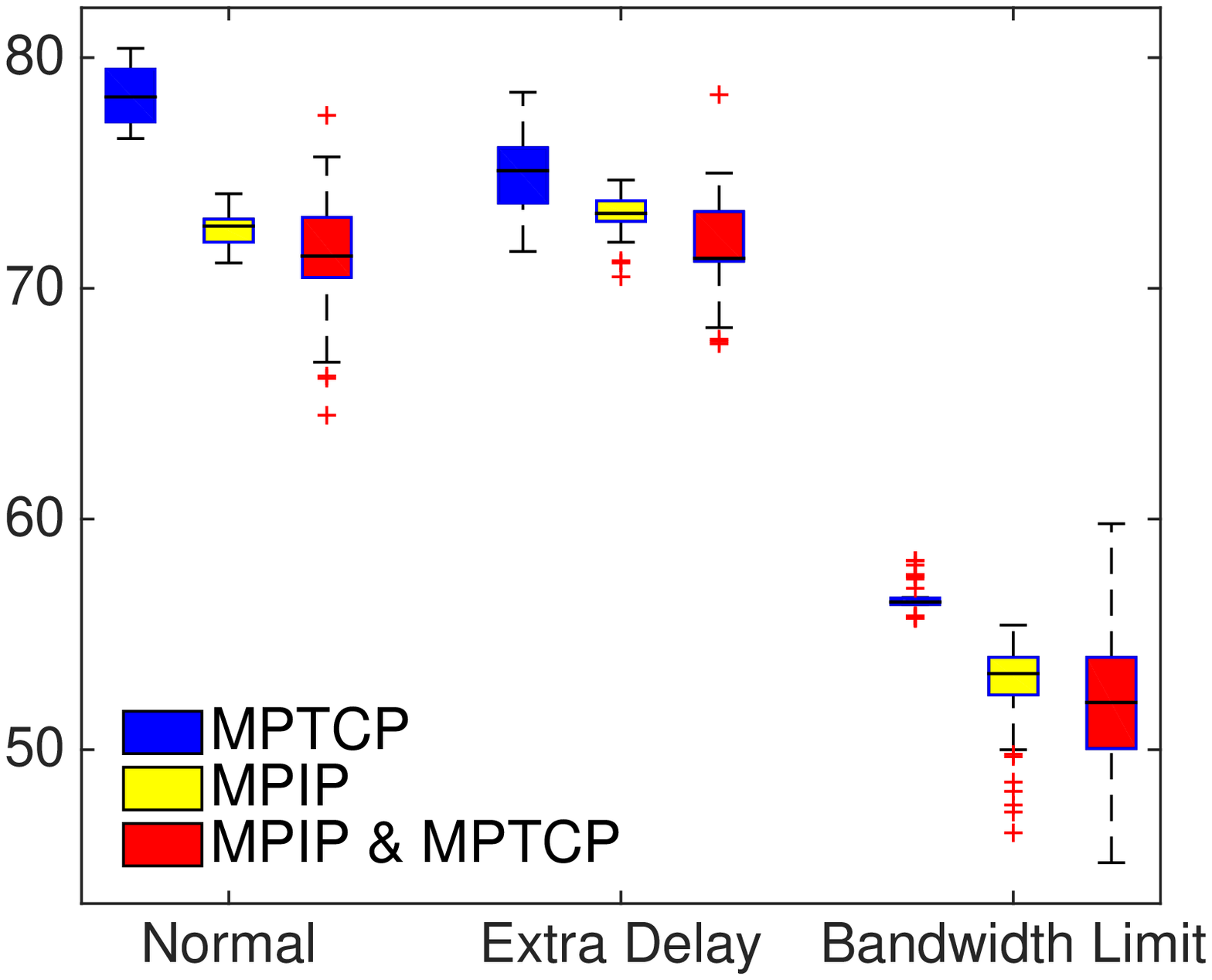}}
\caption{TCP over MPIP Performance}
\label{fig.mpip_path}
\end{figure*}

To test the effectiveness of MPIP load-balancing, we enable only two parallel paths between the two desktops so that they don't share any NIC to prevent traffic coupling. To make it more intuitive, we limit the bandwidth of path $1$ to $40$Mbps and path $2$ to $20$Mbps. From the throughput trend in Figure~\ref{fig.twopathstpcomp}, both paths converged close to their capacities and remained stable for the whole experiment.

Next we compare path failure response time of TCP over MPIP and MPTCP over IP. In Figure \ref{fig:disconnection}, bandwidth of both paths are set to $40$Mbps. In the middle of the experiment, we disconnect one path by unplugging the cable from one NIC to emulate path failure. Both MPIP and MPTCP shift traffic to the surviving path quickly. However, when we plug in the cable after $40$ seconds, MPTCP always suffers a $10$-$20$ seconds delay to re-establish the subflow at the transport layer. Different from MPTCP, MPIP promptly detects the re-activated NIC at the network layer and establishes a new IP path to ramp up the throughput. 

Multipath transmission is vulnerable to out-of-order packet delivery. To test the effectiveness of MPIP's packet reordering mechanism,  we inject extra $10$ms propagation delay to path $1$ and $2$ms delay to path $2$ through the network emulator {\it tc}. 
In \ref{fig:mpip_reorder}, with MPIP reordering disabled, TCP performs poorly with the average throughput around $50$Mbps, and a large number ($3,353$) of retransmissions are detected. 
To the contrast, when MPIP reordering is enabled, TCP throughput is stable and approaches the aggregate capacity of the two paths, and no packet retransmission is detected.

As mentioned in Section~\ref{sec:together}, MPIP should be compatible with MPTCP. Three groups of experiments are conducted for different combinations of multipath transmission at transport and network layers, namely, MPTCP/IP, TCP/MPIP, and MPTCP/MPIP. For the first group (normal), two available paths with $40$Mbps bandwidth each are configured; for the second group (extra delay), an extra $10$ms delay is added to path $1$; at last, bandwidth of path $1$ is limited to $20$Mbps. In Figure \ref{fig:mpip_mptcp}, the boxplots for throughputs of all combinations are plotted. MPTCP/IP throughput is stable and close to the capacity in all cases. TCP/MPIP and MPTCP/MPIP throughputs are little lower but still close to the capacity. Their throughput variances are close larger than MPTCP. This demonstrates that the interaction between MPIP load balancing and upper layer congestion control needs further study and fine-tuning.

\begin{figure*}[htbp]
\centering{
\subfigure[Video Rate under Path Failure \label{fig:dis_tp}]
{\includegraphics[width=0.24\linewidth,height=1.5in]{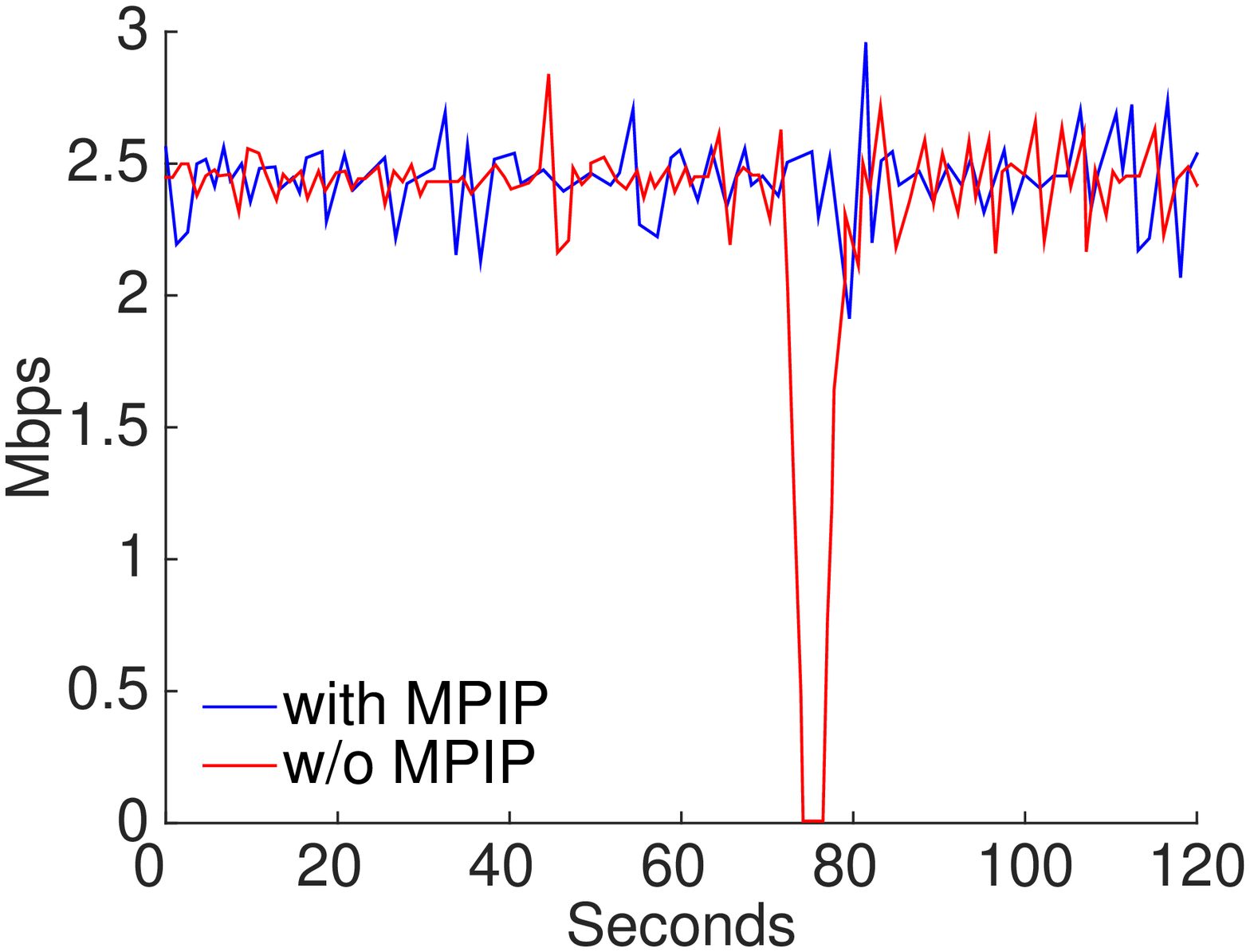}}
\subfigure[Video Rate under Bandwidth Limit \label{fig:limit_tp}]
{\includegraphics[width=0.24\linewidth,height=1.5in]{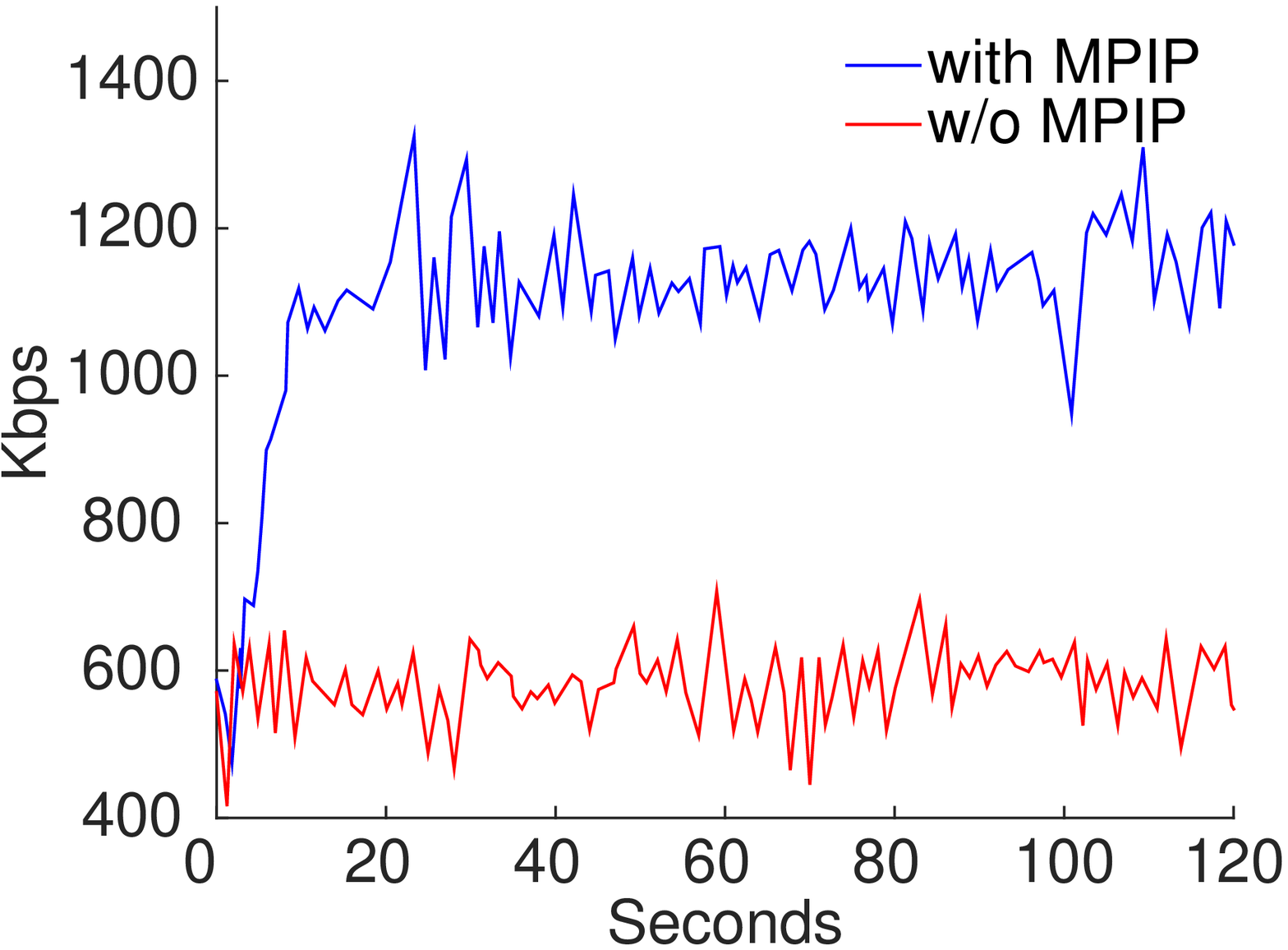}}
\subfigure[WebRTC Audio Delay \label{fig:routing_delay}]{\includegraphics[width=0.24\linewidth,height=1.5in]{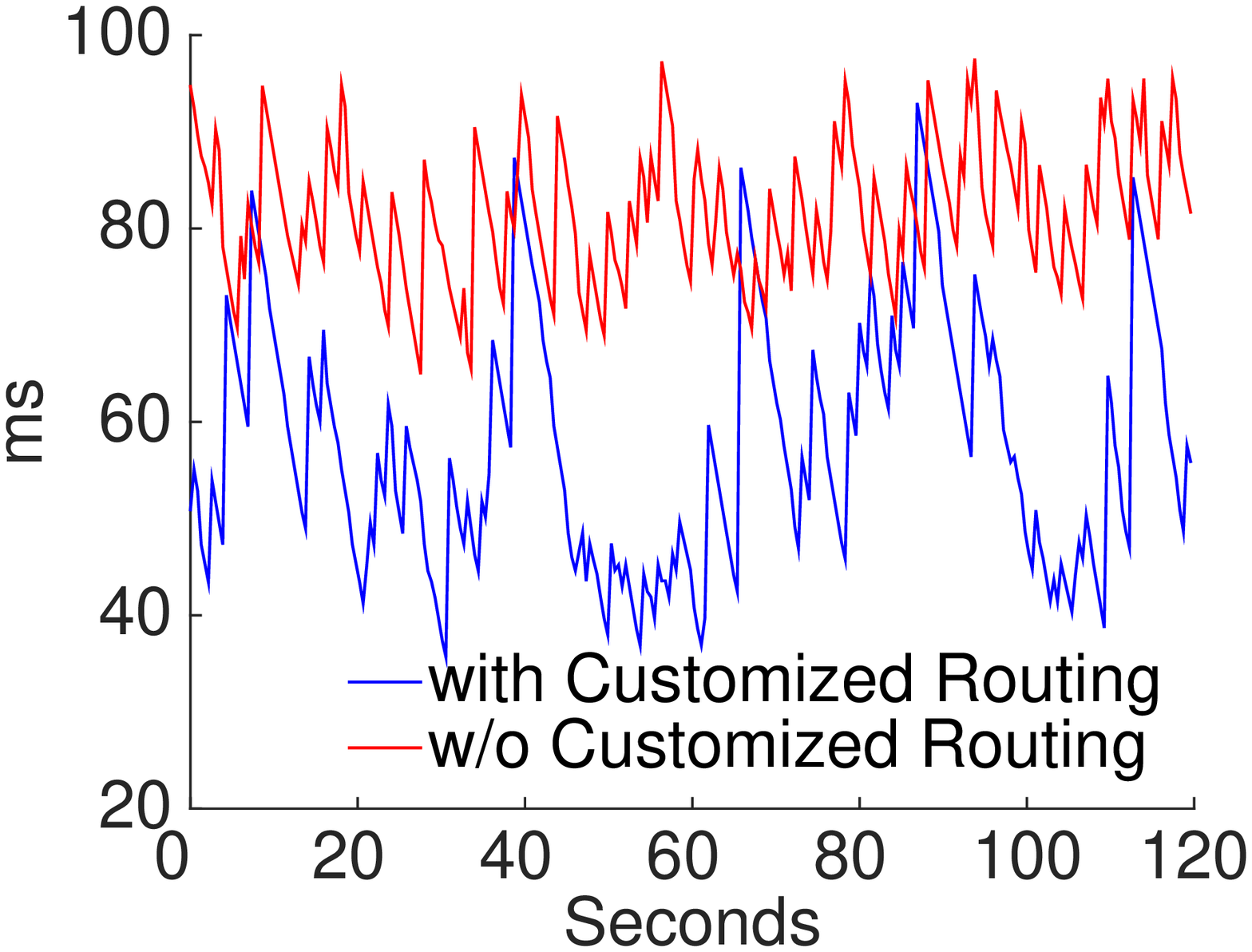}}
\subfigure[WebRTC Video Frame Rate \label{fig:routing_frame}]{\includegraphics[width=0.24\linewidth,height=1.5in]{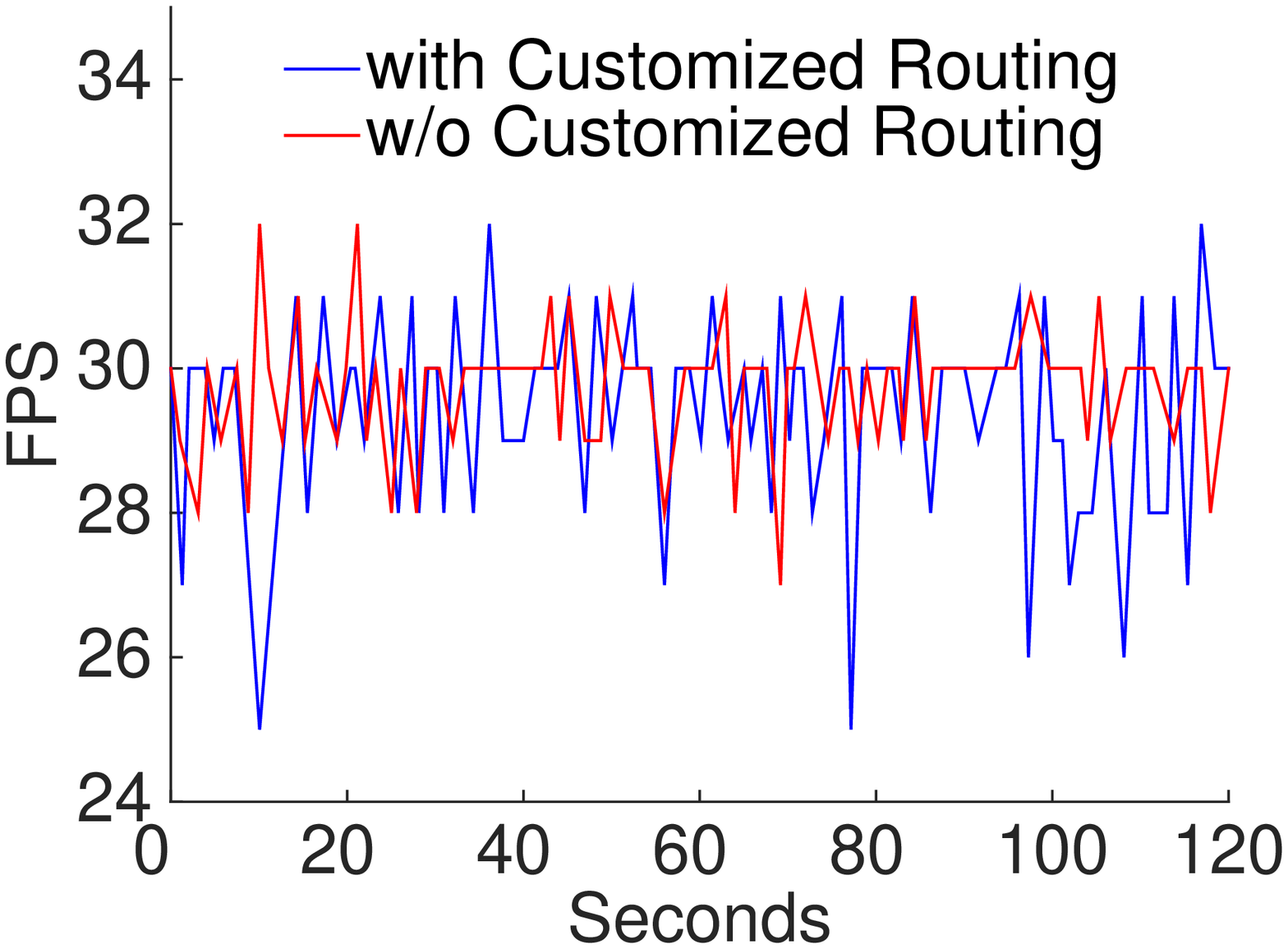}}
\caption{WebRTC Performance over MPIP: (a)(b), all-paths mode; (c)(d), single-path for audio, all-paths for video.} \label{fig:dis_webrtc}}
\end{figure*}

\subsubsection{UDP over MPIP}
To evaluate how UDP-based applications, such as Real Time Communications, can benefit from MPIP, we run WebRTC video chat over MPIP and collect application-level performance by capturing the statistics windows of WebRTC-internals embedded in Chrome, then extracting data from the captured windows using WebPlotDigitizer \cite{rohatgi2011webplotdigitizer}. We first configure two IP paths between two lab machines without bandwidth limit, and then run WebRTC video call between the two machines. To test the robustness of MPIP against path failures, one path is disconnected in the middle of experiment.  As illustrated in Figure \ref{fig:dis_webrtc}, if WebRTC video chat is running over legacy IP, when the original path is disconnected at 72 second, video throughput drops sharply in Figure \ref{fig:dis_tp}, video freezes for few seconds before video flow migrates to the other path. This demonstrates that while WebRTC can recover from path failure at the application layer, its response is too sluggish and user QoE is significantly degraded by a few seconds freezing. With MPIP, video streams continuously without interruption. In addition, to demonstrate how WebRTC benefits from MPIP multipath throughput gain, we limit the bandwidth of each path to $1$Mbps. Comparison presented in Figure \ref{fig:limit_tp} illustrates that with the help of MPIP, WebRTC video throughput improves from $600$Kbps to $1200$Kbps. We then introduce additional delays of $50$ms and $80$ms to the two paths respectively. MPIP then use {\it single-path} mode to route audio packets to the path with shorter delay, while video packets are routed using {\it all-paths} mode. Figure \ref{fig:routing_delay} shows clearly that audio delay is reduced by $30$ms while the video quality is not affected as illustrated in Figure \ref{fig:routing_frame}.



\subsection{Internet Experiments}
\label{sec:lab1}
Besides the controlled lab experiments, we also conduct experiments on the Internet to evaluate MPIP's compatibility with real applications and various middle boxes, e.g. NAT routers, inside ISP and CSP networks.  
\begin{figure}
	\centering
		\includegraphics[width=0.8\linewidth]{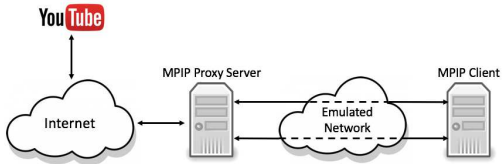}
		\caption{MPIP works with YouTube  through Proxy}
		\label{fig:proxy}
\end{figure}

\begin{figure*}[htbp]
\centering{
\subfigure[Video Buffer Health \label{fig:auto_buffer}]
{\includegraphics[width=0.23\linewidth,height=1.4in]{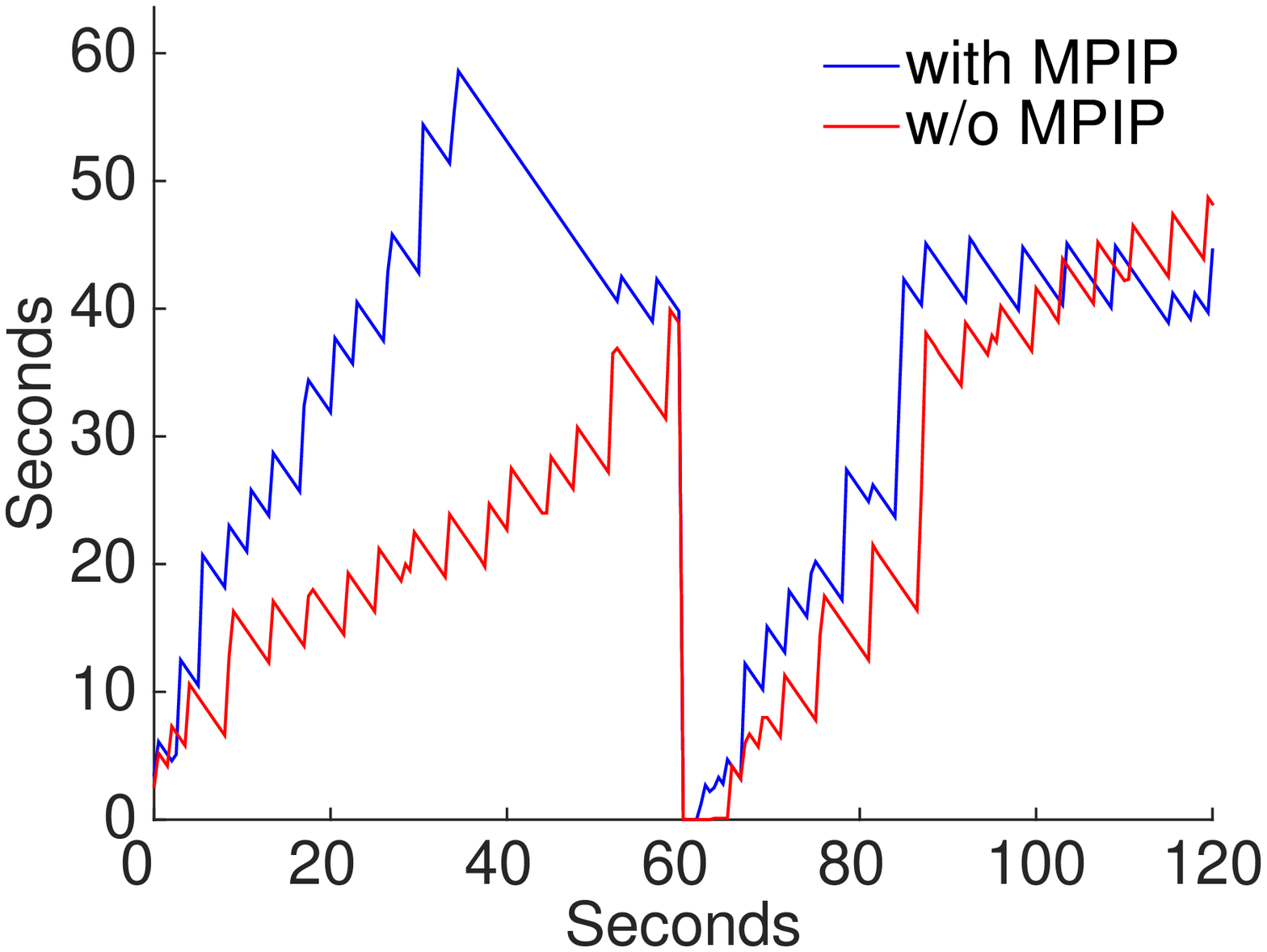}}
\subfigure[Video Resolution\label{fig:auto_resolution}]
{\includegraphics[width=0.23\linewidth,height=1.4in]{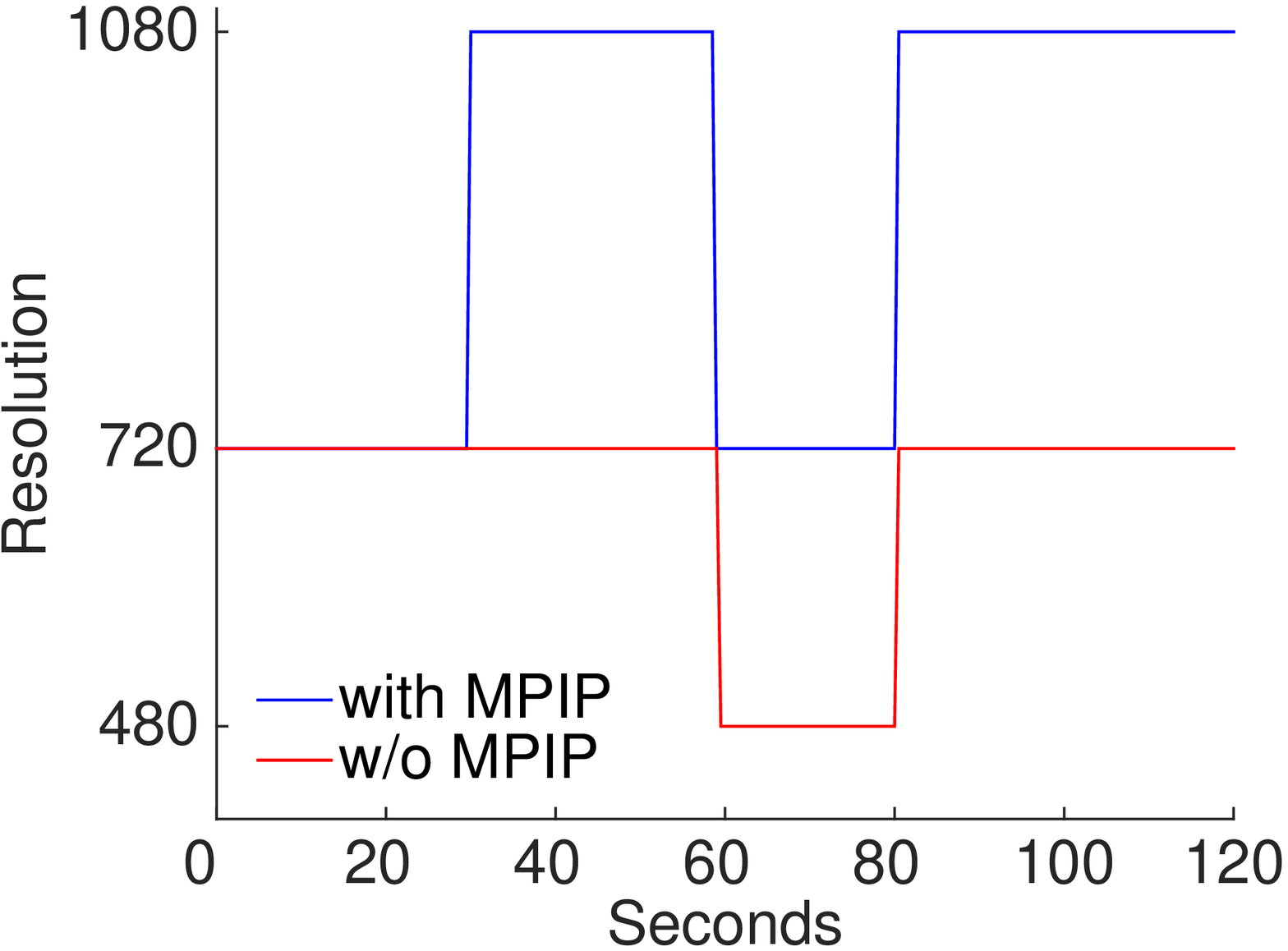}}
\subfigure[Video Frame Rate \label{fig:auto_frame}]
{\includegraphics[width=0.23\linewidth,height=1.4in]{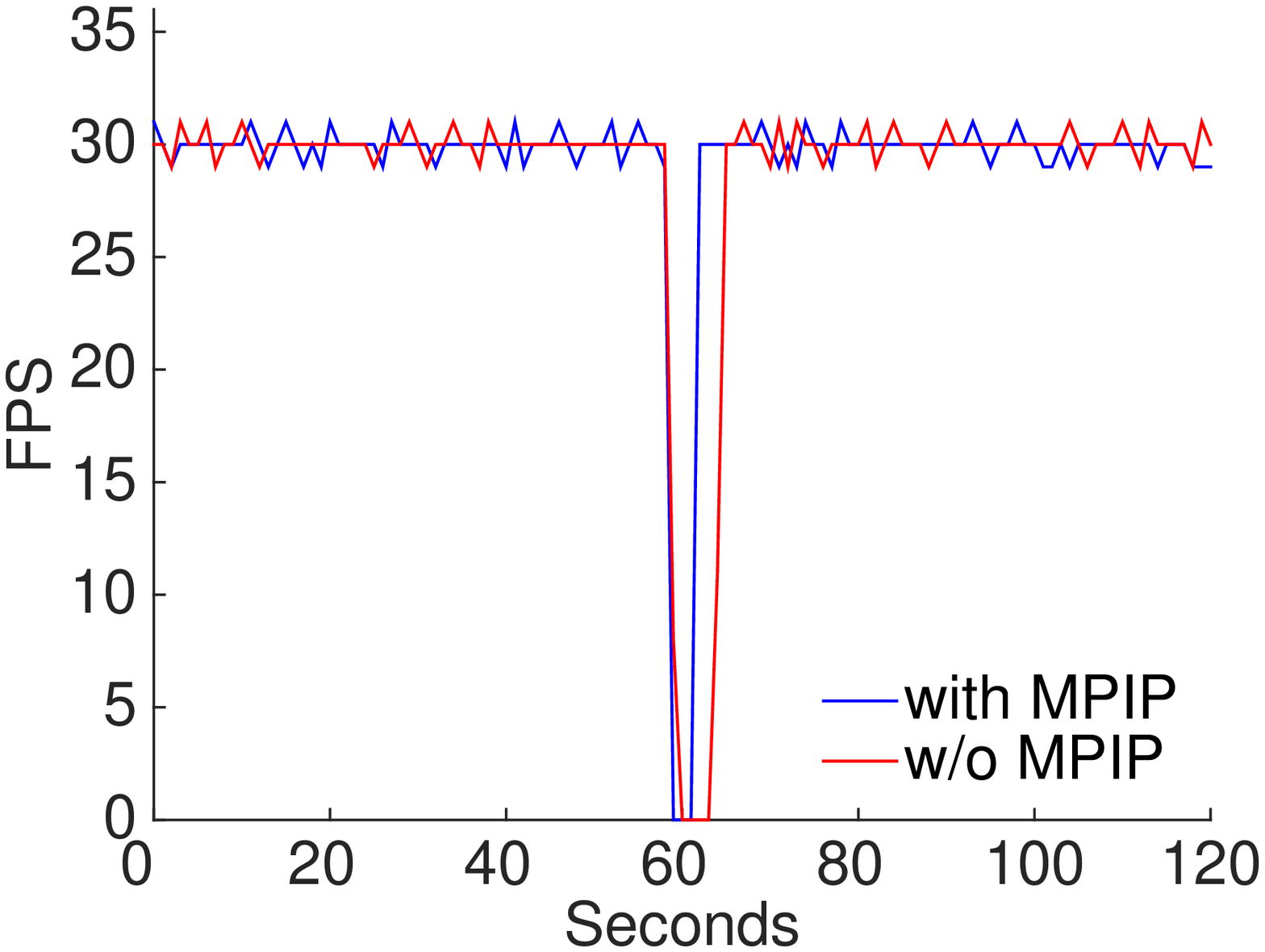}}
}
\subfigure[Video Throughput\label{fig:auto_network}]
{\includegraphics[width=0.23\linewidth,height=1.4in]{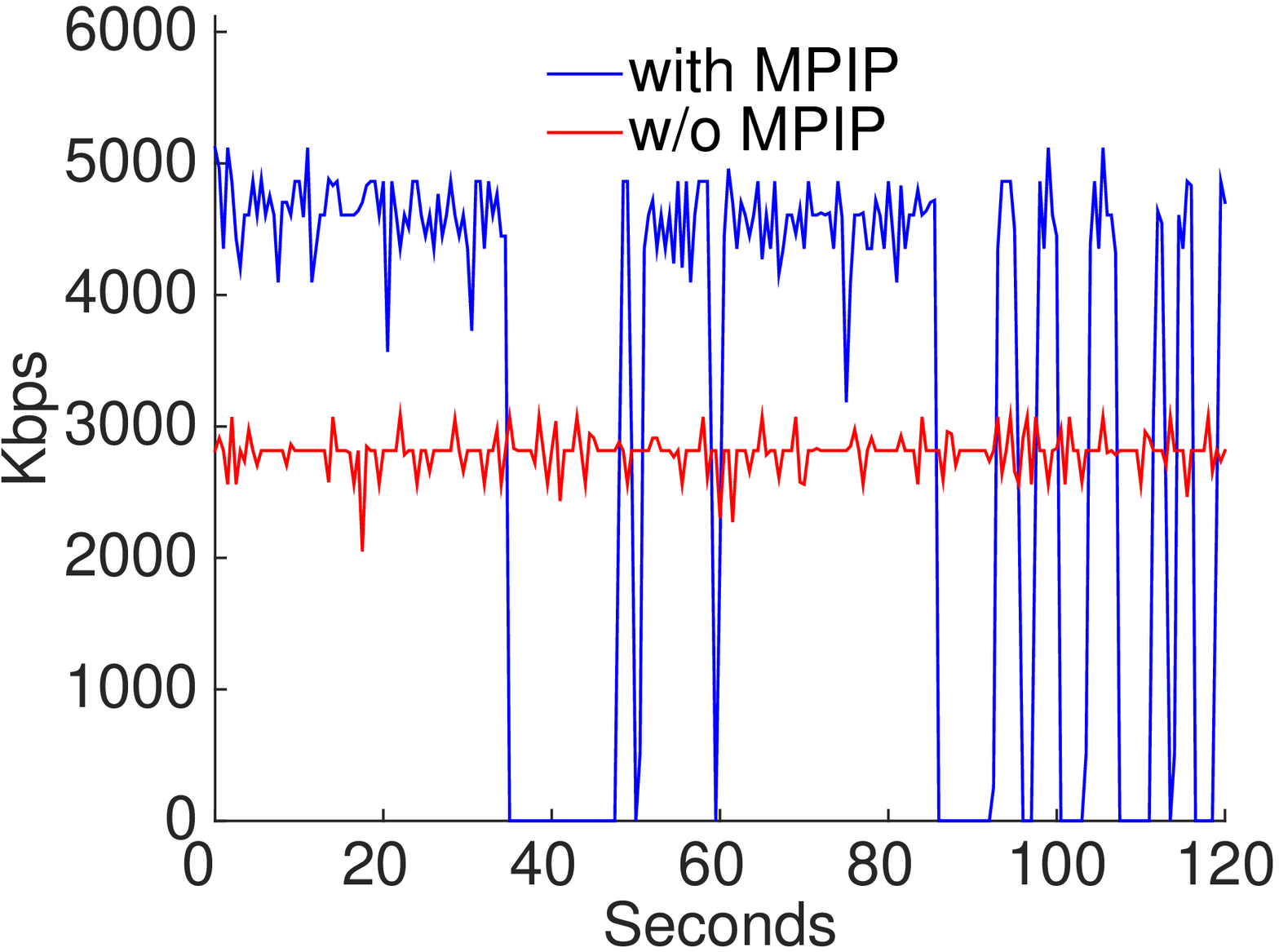}}
\caption{Youtube Video Adaptation with MPIP Client} \label{fig:auto_youtube}
\end{figure*}
\begin{figure*}[htbp]
\centering{
\subfigure[Video Buffer Health \label{fig:720_buffer}]
{\includegraphics[width=0.32\linewidth,height=1.5in]{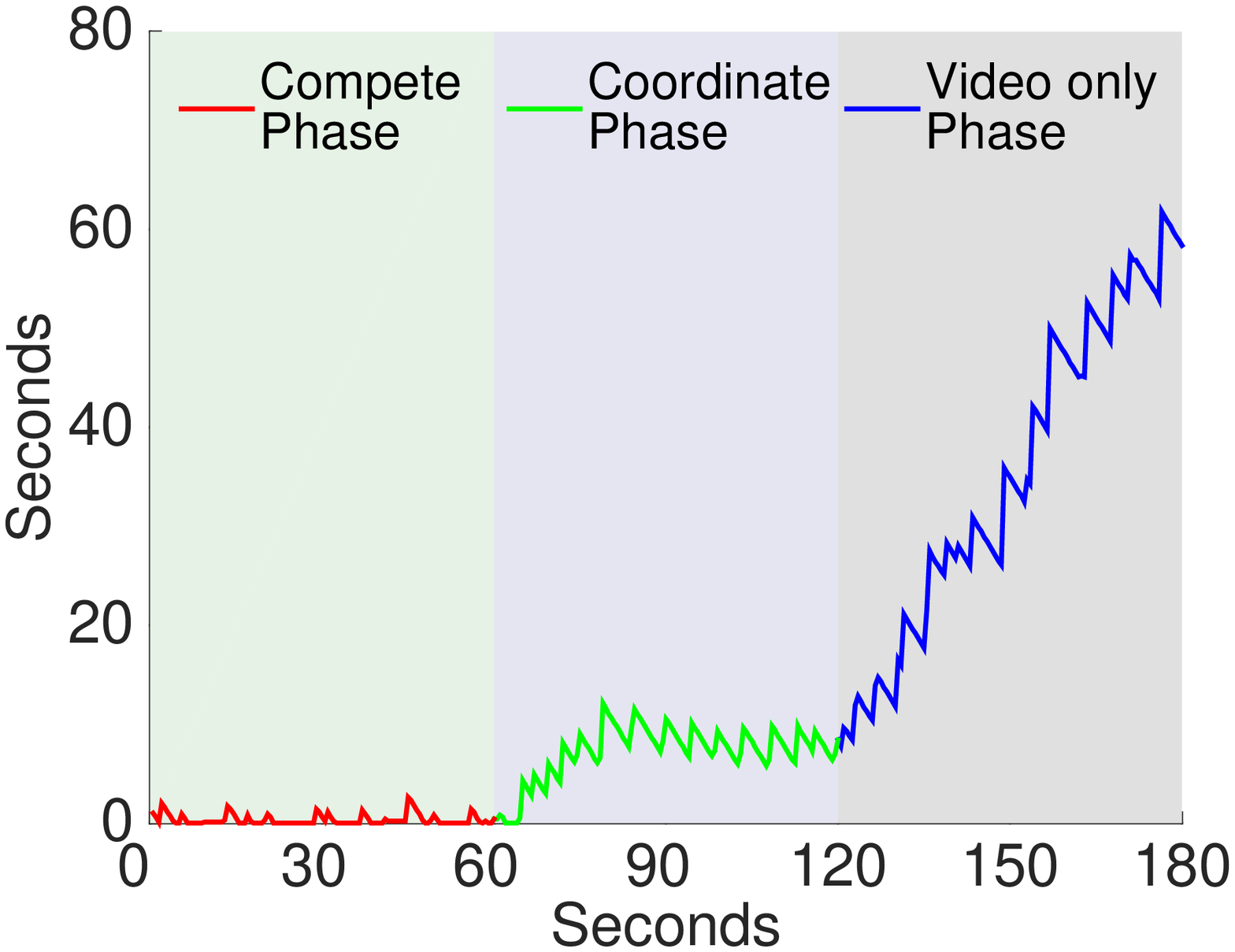}}
\subfigure[Video Throughput\label{fig:720_network}]
{\includegraphics[width=0.32\linewidth,height=1.5in]{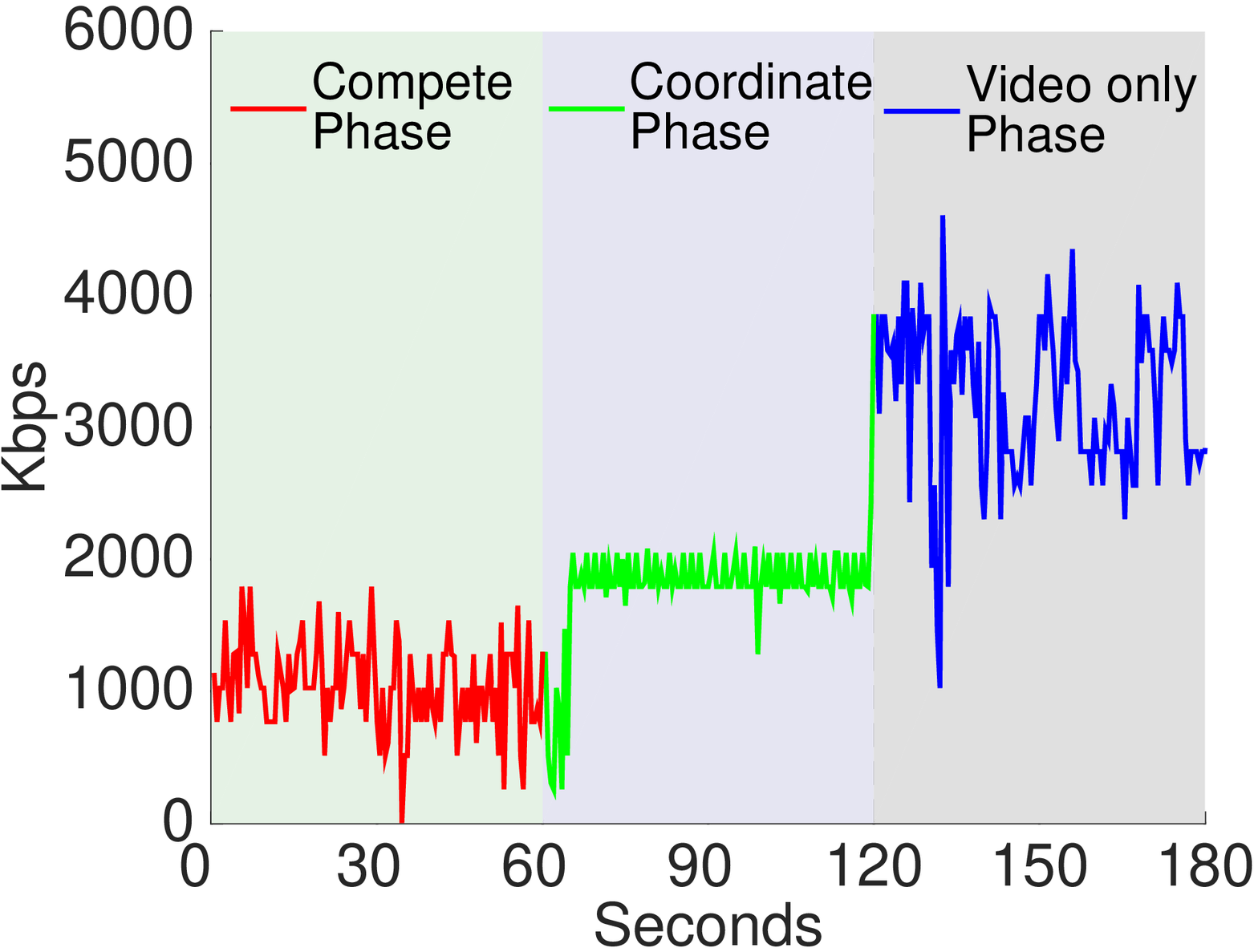}}
\subfigure[Video Frame Rate \label{fig:720_frame}]
{\includegraphics[width=0.32\linewidth,height=1.5in]{{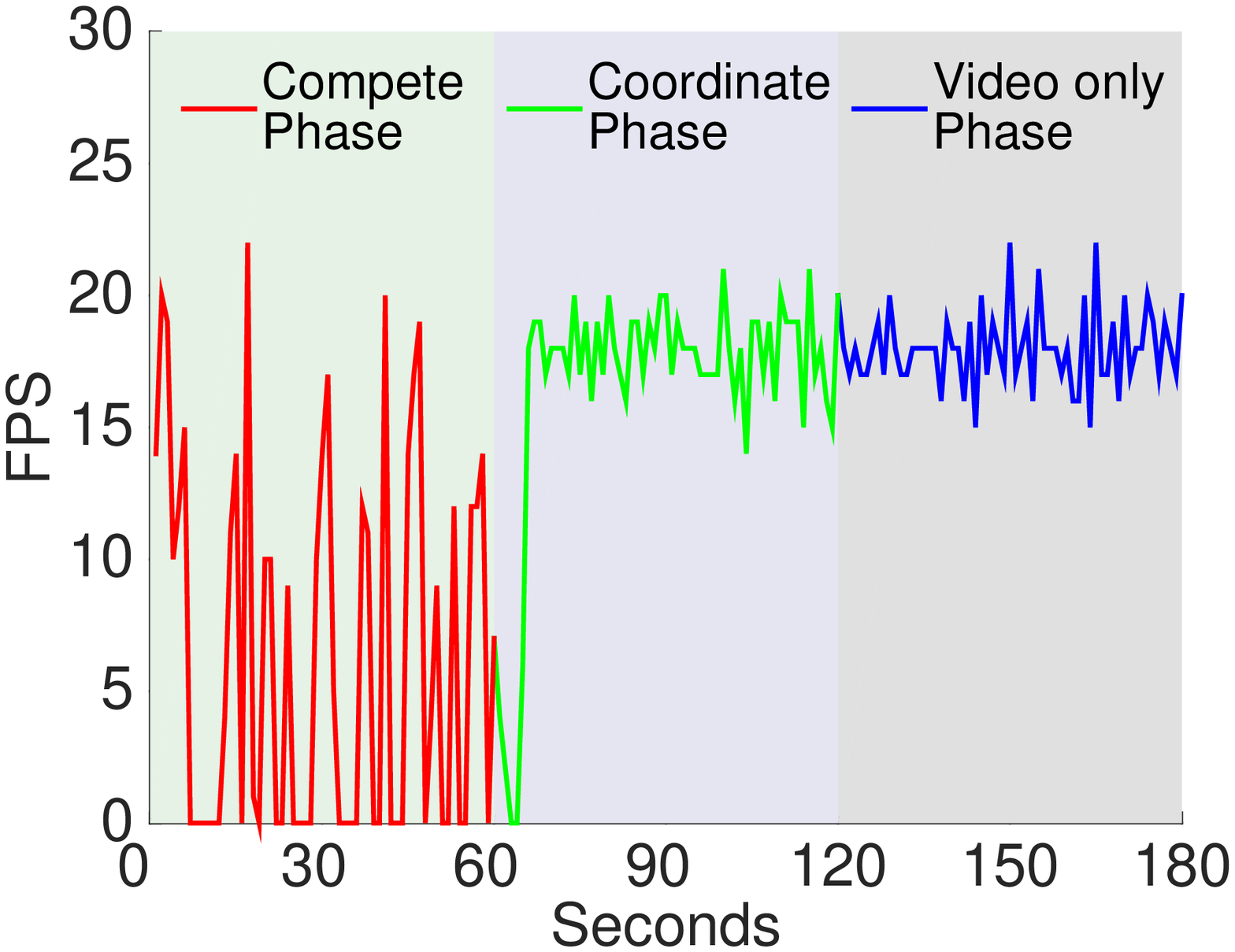}}
}}
\caption{Youtube 720p Video performance with Application Coordination} \label{fig:720_youtube}
\end{figure*}

\subsubsection{YouTube Video Streaming}
Firstly, we measured YouTube video performance to see whether video streaming applications can benefit from MPIP.  Since it's not easy to install MPIP on YouTube servers, we configure a MPIP proxy using Squid on Ubuntu. Three NICs are installed on the proxy server. As illustrated in Figure \ref{fig:proxy}, one NIC is connected to Internet, and the other two are connected to a MPIP client machine with two paths in an emulated network.  
 
While YouTube can run over Google's Quick UDP Internet Connections (QUIC) protocol, but the Squid proxy does not support QUIC. In our experiments, YouTube still runs over HTTP/TCP. The minimum throughout required to stream a $720$P resolution video is no less than $2.5$ Mbps and that for $1080$P is $4.5$Mbps.
%
We limit the bandwidth of each path to $3$Mbps so that the aggregate capacity is sufficient for YouTube $1080$P video. We enable YouTube adaptive streaming where video quality is determined automatically based on the network condition. Figure \ref{fig:auto_resolution} shows that initial video resolution will always be configured to $720$P even if MPIP is running to provide $6$Mbps aggregation bandwidth. However, with MPIP, YouTube can quickly build up the video preload buffer, see \ref{fig:auto_buffer}. When the preload buffer length exceeds $50$ seconds, YouTube increases video resolution from $720$P to $1080$P around $30$ seconds into the experiment. At the $60$th second, we manually fast-forward the video outside the preload buffer coverage, then video resolution in both cases drop one level down and recover back to the previous level about $20$ seconds later. Video frame rate can be sustained at $30$FPS except when the fast-forward happens (Figure~\ref{fig:auto_frame}). And as long as the preload buffer length goes over $40$ seconds, YouTube video client will pause video chunk request from the server. This explains frequent MPIP throughput 
dips in Figure~\ref{fig:auto_network}. Video playback is smooth due to preload buffer. 
 



\subsubsection{Coordination between Applications}
Applications running on the same machine compete for network resources. In this part, we demonstrate that MPIP can select paths for applications in a coordinated fashion to maximize the aggregate performance. We reused the testbed in Figure~\ref{fig:proxy} with $2$Mbps bandwidth limit for each path and introduced $20$ms extra delay to one path. Experiments were conducted in three phases with MPIP enabled all the time. At the beginning, besides the YouTube video session, another file downloading session is added to transmit data from MPIP proxy server to client. Initially MPIP operates in the {\it all-paths} mode and establishes two paths for each session to acquire more bandwidth. Due to the path delay difference, out-of-order packet deliveries limit the TCP throughput for both sessions. Sixty seconds into the experiment, MPIP applies coordinated routing for the two sessions: both sessions are routed using the {\it single-path} mode, with the video session assigned to the path with shorter delay and the file downloading session assigned to the other path. As illustrated in Figure \ref{fig:720_youtube}, coordinated routing significantly improve the performance of the video session:   video throughout increases by $400$Kbps (from $1,500$Kbps to $1,900$Kbps), value of FPS stabilizes around $20$  without freezing, and buffer length accumulates to $10$ seconds. Meanwhile, the average throughput of the downloading session drops from $2.51$Mbps to $1.89$Mbps. Since users are more sensitive to video quality than the file downloading throughput, the coordinated routing presumably improves the overall user experience. Sixty seconds later, we terminated the downloading session. From Figure \ref{fig:720_buffer}, \ref{fig:720_network} and \ref{fig:720_frame}, we observe that both the video throughput and preload buffer length increase significantly, while FPS of the video doesn't change much. Longer preload buffer length leads to more stable video playback.
\\


\begin{figure}[htbp]
\centering{
\subfigure[WiFi and Cellular \label{fig:wifi_cell}]
{\includegraphics[width=0.49\linewidth,height=1.3in]{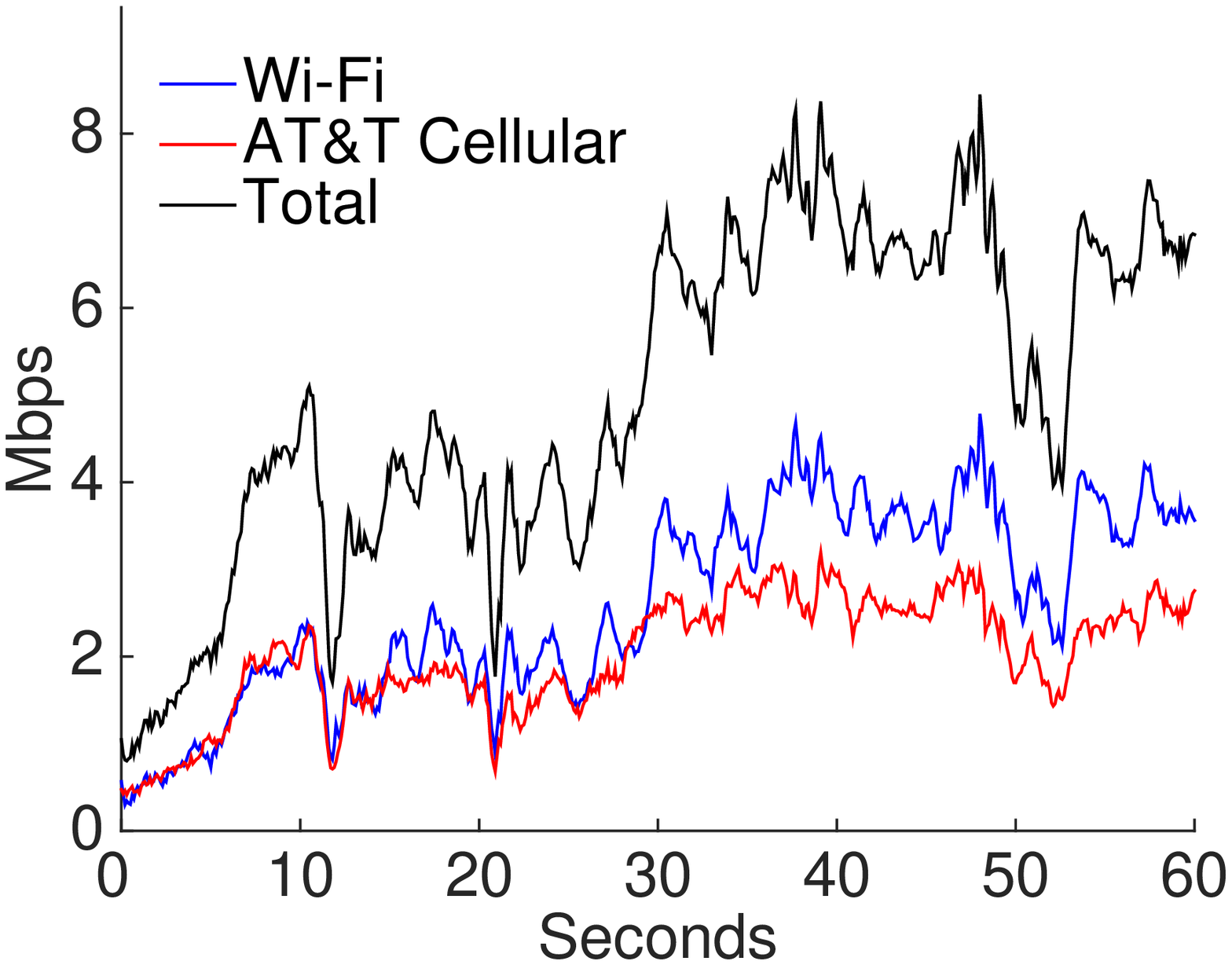}}
\subfigure[ Two Cellular Networks \label{fig:cell_cell}]
{\includegraphics[width=0.49\linewidth,height=1.3in]{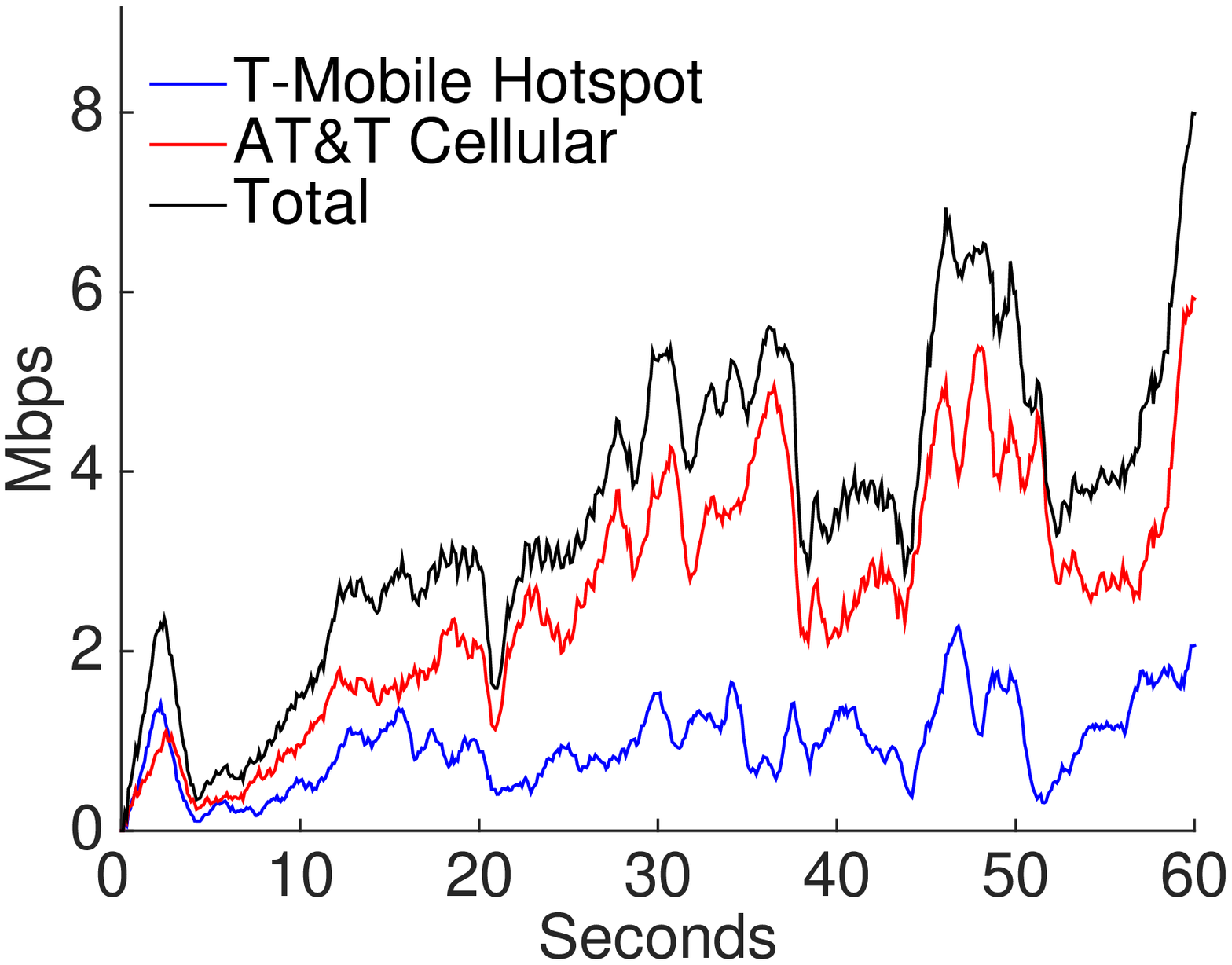}}}
\caption{MPIP over Wireless} \label{fig:mpip_real_network}
\end{figure}


%
\subsection{Android Experiments}
We use a Nexus $5$X phone located in California to test Android MPIP. The phone is equipped with one cellular interface and one WiFi interface. We use it to download data from a server located in New York City with one public IP address. We first connect the phone to a corporate ISP through WiFi and AT\&T CSP through 4G cellular. Without MPIP, the phone can achieve  average bandwidth of $4.5$Mbps through WiFi and $4.3$Mbps through cellular respectively. The average RTTs of WiFi and cellular are $76.2$ms and $155.9$ms. When MPIP is enabled, as illustrated in Figure \ref{fig:mpip_real_network},  Android MPIP can concurrently transmit data on both paths going through different ISP/CSP and reach aggregate throughput of $7.5$Mbps in the face of large delay disparity. Next we replace the corporate WiFi router with a hotspot hosted by another phone connected to T-Mobile cellular network. As all data through the  hotspot are forwarded by another phone, the average RTT on the T-Mobile path increases dramatically to $349.2$ms and the average bandwidth is only  $1.52$Mbps. Figure \ref{fig:cell_cell} demonstrates that even when one cellular path has bad performance, MPIP still manages to multiplex bandwidth from two CSPs to achieve higher aggregate throughput. 

%% file: related.tex
\vspace{-0.2cm}
\section{Related Work}
\label{sec:related}
Multipath transmission is a fundamental technique for network traffic engineering, e.g., Multiprotocol Label Switching (MPLS). Multipath routing can be used in ad hoc networks to increase data transfer throughout and robustness~\cite{multi_adhoc1, multi_adhoc2, multi_adhoc3}. It is also used in emerging network architecture, e.g.~\cite{multi_ICN}. The growing popularity of multi-homed devices makes it possible to initiate multipath transmission from end devices. Back to $2001$, Hsieh et al proposed pTCP\cite{hsieh01} that effectively performs bandwidth aggregation on multi-homed mobile hosts. In \cite{key01}, the authors investigated the potential benefits of coordinated congestion control for multipath data transfers. In \cite{dong01}, Dong et al implemented concurrent TCP(cTCP) in FreeBSD to improve throughput. Also, the Stream Control Transmission Protocol (SCTP)\cite{sctp, joe01} is an early protocol designed for multihoming to support failover and simultaneous transmission. In $2010$, Barre et al published experimental results of using multiple paths simultaneously in TCP transmission~\cite{BRBH10}. Based on IETF RFC $6182$ for multipath TCP in $2011$, the same team implemented a complete prototype of multipath TCP in Linux and Android system \cite{mptcp}. They also explored many other aspects of MPTCP in \cite{PDDRB12}, \cite{DPLMAB13}, \cite{PKB13}, \cite{PFAB14}. In \cite{chen01}, Chen et al did a thorough measurement of MPTCP over wireless links. In \cite{cao01}, a variation of TCP Vegas~\cite{vegas}, was proposed for multipath TCP. Different from those multipath protocols at the transport layer, MPIP is a transparent 
multipath solution at the network layer of end devices. 

As bandwidth of cellular network becomes comparable with the wired Internet, switching among WiFi and cellular becomes practical for mobile devices. IETF released RFC $5206$\cite{hip} to propose a draft of host identity instead of IP address for mobile devices that have multiple interfaces. In \cite{singh01}, the authors designed a complete system that supports smooth transfer among different networks. Shuo et al proposed a transport framework of mobile network selection named Delphi in \cite{deng01}. Delphi chooses the best path for applications based on network properties. A system named MultiNets proposed in \cite{nirjon01} chooses the best interface on a mobile device based on energy consumption, data usage charge and throughput consideration.  All these solutions require significant changes and coordination at multiple layers. In~\cite{multi_handoff2}, a pure user-level solution, called msocket, was proposed for seamless handover between different mobile networks. Different from these previous work, MPIP realizes path selection and seamless handover by only changing the network layer. It has long been observed that routing for applications on the same device needs to be coordinated~\cite{congestion_manager1,congestion_manager2}. MPIP serves as a light-weight framework to facilitate coordinated routing for multiple applications over multiple paths. The stability, fairness and efficiency of multipath protocols have been studied in different contexts~\cite{han2004overlay, peng2014multipath,khalili2013mptcp,multipath,multipath1}. 